\documentclass{ametsocV6.1}
\nolinenumbers

\newcommand{\bx}{\mathbf{x}}
\newcommand{\by}{\mathbf{y}}

\newcommand{\bv}{\mathbf{v}}

\newcommand{\bH}{\mathbf{H}}
\newcommand{\bR}{\mathbf{R}}
\newcommand{\bB}{\mathbf{B}}

\newcommand{\bP}{\mathbf{P}}
\newcommand{\bK}{\mathbf{K}}

\newcommand{\bI}{\mathbf{I}}
\newcommand{\bO}{\mathbf{0}}

\newcommand{\beps}{\pmb{\varepsilon}}

\title{Using Diffusion Models to do Data Assimilation}

\authors{
	Daniel Hodyss,\aff{a}\correspondingauthor{Dr. Daniel Hodyss, daniel.h.hodyss.civ@us.navy.mil}
    Matthias Morzfeld\aff{b}
}

\affiliation{
	\aff{a}{Remote Sensing Division, U.S. Naval Research Laboratory, Washington DC}\\
    \aff{b}{Cecil H. and Ida M. Green Institute of Geophysics and Planetary Physics, Scripps Institution of Oceanography, University of California, San Diego, CA}
}

\abstract{
\nolinenumbers
The recent surge in machine learning (ML) methods for geophysical modeling has raised the question of how these methods might be applied to data assimilation (DA). We focus on diffusion modeling (a form of generative artificial intelligence) for systems that can perform the entire DA process, rather than on ML-based tools used within a conventional DA system. We identify at least three distinct types of diffusion-based DA systems and show that they differ in the posterior distribution they target for sampling. These posterior distributions correspond to different priors and/or likelihoods, which in turn result in unique training datasets, computational requirements, and state estimate qualities. Our analysis further shows that a diffusion DA system designed to target the same posterior distribution as current ensemble DA algorithms requires re-training at each DA cycle, which is computationally costly. We discuss the implications of these findings for the use of diffusion modeling in DA.}

\begin{document}

%% Necessary!
\maketitle

%%%%%%%%%%%%%%%%%%%%%%%%%%%%%%%%%%%%%%%%%%%%%%%%%%%%%%%%%%%%%%%%%%%%%
% SIGNIFICANCE STATEMENT/CAPSULE SUMMARY
%%%%%%%%%%%%%%%%%%%%%%%%%%%%%%%%%%%%%%%%%%%%%%%%%%%%%%%%%%%%%%%%%%%%%
%
% If you are including an optional significance statement for a journal article or a required capsule summary for BAMS 
% (see www.ametsoc.org/ams/index.cfm/publications/authors/journal-and-bams-authors/formatting-and-manuscript-components for details), 
% please apply the necessary command as shown below:
%
% Significance Statement (all journals except BAMS)
%
%\statement
%	 Enter significance statement here, no more than 120 words. See \url{www.ametsoc.org/index.cfm/ams/publications/author-information/significance-statements/} for details.
%
%% Capsule (BAMS only)
%%
%\capsule
%       Enter BAMS capsule here, no more than 30 words. See \url{www.ametsoc.org/index.cfm/ams/publications/author-information/formatting-and-manuscript-components/#capsule} for details.
%
%% * * If using twocol mode, you will need to use the commands "twocolsig" and "twocolcapsule" in place of "sig" and "capsule"
%%      to ensure that the text box correctly spans across both columns.
%

%%%%%%%%%%%%%%%%%%%%%%%%%%%%%%%%%%%%%%%%%%%%%%%%%%%%%%%%%%%%%%%%%%%%%
% MAIN BODY OF PAPER
%%%%%%%%%%%%%%%%%%%%%%%%%%%%%%%%%%%%%%%%%%%%%%%%%%%%%%%%%%%%%%%%%%%%%

\section{Introduction}
Data assimilation (DA) is a mathematical and computational framework for updating forecasts in view of observations \citep[see, e.g.,][]{Kalnay_2002}. 
Mathematically, DA relies on Bayes' rule and all numerical methods for DA can be understood as approximating, in one way or another, a Bayesian posterior distribution.
In numerical weather prediction (NWP), we distinguish between ensemble DA and variational methods.
Ensemble DA includes Monte Carlo based, ensemble Kalman methods \citep[see, e.g.,][]{E94,EAKF,TetAl03,E09Book}, and ``hybrid'' methods that combine the Monte Carlo approach with optimization \citep[see, e.g.,][]{HS00, L03, ZZH09, BMC13, KRBMB13, PZ15}.
Traditional variational DA (3D- or 4D-Var) uses optimization to find optimal state estimates, but does not randomly sample from a posterior distribution to form an ensemble \citep[see, e.g.,][]{T87}.
Collectively, we will refer to ensemble and variational DA as ``conventional DA,'' because these methods have been deployed operationally for the past few decades.

Recently, there has been a surge in interest in applying machine learning (ML) methods to geophysical modeling \cite[see, e.g.,][]{gencast,graphcast,neuralGCM}, which raises the question of how ML tools might be used for DA.
A first step might be to replace the physics-based forecast model within a conventional DA system with a data-driven ML emulator while continuing to employ conventional DA methods to merge forecasts with observations \citep[see, e.g.,][]{ASAW25}.
A more ambitious step might be to replace the entire conventional DA system with ML.
This form of ML-based DA system can be implemented via diffusion models
\citep[a form of generative artificial intelligence (AI), see, e.g.,][]{rozet2023,simpleObLike2024, manshausen2024,pathak2024, Qu2024, li2025}, deep neural networks \citep[see, e.g.,][]{Aardvark} or by training a neural network that represents the functional of variational data assimilation, even in the absence of a training dataset \citep{KP24}.

This paper explores the question: \emph{Can a diffusion model completely replace a conventional DA system?}  To answer this, we must first understand the subtle differences in the Bayesian posterior distributions that arise from different assumptions about the available prior distribution, which, in an ML/AI context, refers to the training set. An ensemble-based conventional DA system samples a Bayesian posterior distribution with a time-dependent prior that is informed by past observations. We will refer to this as a \emph{cycling prior} because it propagates information from previous DA cycles to the current one.  In contrast, we will demonstrate that a diffusion model that is trained on a long time-series of past weather data targets a different Bayesian posterior distribution with a constant, \emph{climatological prior}. While some diffusion DA systems incorporate a ``forecast" of some sort along with observations, we will show that this is still not the same as a cycling prior.  Ultimately, we conclude that, except in rare cases (\cite{BAO2024}), ensemble DA and diffusion DA target different posterior distributions. We will formulate our arguments using a simple linear example with Gaussian errors, which allows for a clear and exact analysis without approximations.  This leads to a crucial question: \textit{Which diffusion model formulation minimizes variance and forecast errors?} Our analysis will unambiguously show that a diffusion model designed to sample the posterior distribution with a cycling prior, which is the one also targeted by current ensemble DA, is the optimal choice.

On the other hand, a DA method's practical performance depends on how accurately it can sample the posterior distribution. In nonlinear systems, these distributions are often non-Gaussian, causing conventional DA methods that assume Gaussian distributions to produce significant errors (\cite{MH19}). This difficulty in sampling the posterior can obscure the theoretical differences between various DA methods.  While important, analyzing these application-dependent sampling issues is beyond the scope of this paper. Instead, we focus on a more fundamental question: \textit{What specific variations of Bayesian posterior distributions are the various diffusion models sampling?} By identifying this, we can clearly distinguish the algorithms in a way that is valid for all systems, whether they are linear, nonlinear, Gaussian, or non-Gaussian. The practical impact of non-Gaussianity is left for future work.

The rest of this paper is organized as follows.
In Section 2 we will introduce ensemble and variational methods as conventional DA systems.  In Section 3 we introduce diffusion modeling and diffusion DA. In Section 4 we describe a linear, stochastic dynamical system that will allow us to clearly formulate all aspects of the DA problem analytically.  We will apply different forms of diffusion DA systems to this dynamical model in order to reveal the prior, likelihood and posterior each system corresponds to. In Section 5 we provide a numerical illustration of the main results from Section 4. In Section 6 we provide a non-technical overview of the main results from sections 4 and 5.  We close the manuscript with a summary of the major results and their conclusions in Section 7.

\section{Ensemble and variational data assimilation}
\label{sec:conv-diffy}
We briefly review fundamental concepts of ensemble and variational DA with an emphasis on how the methods target different Bayesian posterior distributions that are generated by different prior assumptions. 

\subsection{Ensemble data assimilation}
\label{sec:EnKF}
Ensemble DA is concerned with \emph{sampling} a time-evolving Bayesian posterior distribution that describes the probability of a system state $\bx_k$ at (discrete) time $k$, given observations $\by_1,\by_2,\dots,\by_{k-1},\by_{k}$ up to time $k$:
\begin{equation}
	\label{iterated}
	p(\bx_k\vert \by_1,\dots,\by_k) \propto p(\by_k\vert \bx_k) p(\bx_k\vert \by_1,\dots,\by_{k-1}).
\end{equation}
It is important to note here that information is propagated from cycle-to-cycle by the time evolving, or ``cycling'' prior $p(\bx_k\vert \by_1,\dots,\by_{k-1})$. In other words, the posterior of the previous cycle is used to generate the prior for the next cycle.  

The ensemble Kalman filter (EnKF) generates ensembles that are approximately distributed according to the posterior distribution~\eqref{iterated} as follows  \citep[see, e.g.,][]{BvLE98, E94,E09Book}.  
At cycle $k-1$, we have observations $\by_1,\dots,\by_{k-1}$ and samples from $p(\bx_{k-1}\vert \by_1,\dots,\by_{k-1})$ in the form of an ensemble $\bx_{k-1}^{(i)}$,
where superscript $i=1,\dots,n_e$ indexes the ensemble members and where $n_e$ is the ensemble size.
The EnKF makes a \emph{forecast} for time $k$ by evolving the ensemble $\bx_{k-1}^{(i)}$ forward in time using the model;
the result is a forecast ensemble, viz.
\begin{equation}
\label{eq:Model}
	\bx_{fk}^{(i)} = \mathcal{M}(\bx_{k-1}^{(i)}),
\end{equation}
where $\mathcal{M}(\cdot)$ is a (possibly stochastic) forecast model, either physics-based or data-driven ML/AI.
This forecast ensemble represents the cycling prior at time $k$, $p(\bx_k\vert \by_1,\dots,\by_{k-1})$, as a collection of samples.
The forecast ensemble is updated to an \emph{analysis} ensemble by employing stochastic ensemble generation (\cite{vanLeeuwen}), viz.
\begin{equation}
	\label{eq:EnKFUpdate}
	\mathbf{x}_{ak}^{(i)} = \mathbf{x}_{fk}^{(i)} + \mathbf{K}(\mathbf{y}_{k}-(\mathbf{H} \mathbf{x}_{fk}^{(i)}+\boldsymbol \varepsilon^{(i)})),
\end{equation} 
where we assume for ease of presentation that the observation operator is linear, i.e.,
\begin{equation}
	\label{eq:ObsEq}
	\mathbf{y}_k = \bH \bx_k + \boldsymbol \varepsilon,
\end{equation} 
where $\boldsymbol \varepsilon$ is a Gaussian random variable with mean $\mathbf{0}$ and covariance matrix $\bR$;
$\boldsymbol \varepsilon^{(i)}$ in~\eqref{eq:EnKFUpdate} is the $i^{th}$ sample from the random variable $\boldsymbol \varepsilon$;
the matrix 
\begin{equation}
	\bK = \bP \bH^T (\bH \bP \bH^T +\bR)^{-1},
\end{equation}
is an ensemble approximation of the Kalman gain and $\bP$ is the ensemble covariance,
usually localized and inflated \citep[see, e.g.,][]{MH23,Whitaker2012,HCW16,Gharamti2019}). 

There also exist hybridized methods between the EnKF presented above and variational methods.  Hybrid variational methods combine gradient-based optimization with an ensemble approach and target a similar Bayesian posterior distribution, but now randomly sample the state \emph{before} the observation time (this is typically referred to as a ``smoother''; see \cite{HBM2016} for a review):
\begin{equation}
	p_\text{EDA}(\bx_{k-1}\vert \by_1,\dots,\by_k) \propto p(\bx_{k-1}\vert \by_1,\dots,\by_{k-1})  p(\by_k\vert \bx_{k-1}),
\end{equation}
where $p(\bx_{k-1}\vert \by_1,\dots,\by_{k-1})$ is a cycling prior and $p(\by_k|\bx_{k-1})$ is the likelihood, connecting the state at time $k-1$  to the observation at time $k$:
\begin{equation}
	\by_k = \bH \mathcal{M}(\bx_{k-1}) + \boldsymbol \varepsilon.
\end{equation}
Specifically, an ``ensemble of data assimilation's'' \citep[EDA,][]{BIH12} maximizes randomized posterior distributions that are obtained by assuming that the cycling prior $p(\bx_{k-1}\vert \by_1,\dots,\by_{k-1})$ is Gaussian with mean $\boldsymbol \mu_{k-1}$ and flow-dependent covariance matrix $\bB_{k-1}$, both generated from the ensemble at the previous cycle. The randomized negative log of the posterior distribution defines the loss function
\begin{equation}
    \label{eq:EDA}
	\mathcal{L}_\text{EDA}(\bx_{k-1}) = \frac{1}{2}\left\| \bR^{-1/2} (\by_k-\bH\mathcal{M}(\bx_{k-1}) + \boldsymbol \varepsilon^{(i)} \right\|^2 + 
	\frac{1}{2}\left\| \bB_{k-1}^{-1/2} (\bx_{k-1}-(\boldsymbol \mu_{k-1}  + \boldsymbol \eta^{(i)})) \right\|^2,
\end{equation}
where $\boldsymbol \varepsilon^{(i)}$ is Gaussian with mean zero and covariance matrix $\bR$ and $\eta^{(i)}$ is the $i^{th}$ ensemble perturbation around the ensemble mean $\mu_{k-1}$.  The minimum of the loss in equation (\ref{eq:EDA}) is taken as the $i^{th}$ posterior ensemble member, which in the linear, Gaussian case reduces precisely to the stochastic ensemble in equation (\ref{eq:EnKFUpdate}).  Thus, the EnKF and the EDA are quite closely related sampling strategies for generating ensembles with similar statistics.

For our purposes of interpreting diffusion DA in the context of Bayesian posterior distributions, the cycling nature of the priors used in the EnKF or EDA is important: The prior at the current cycle is directly informed by previous observations and the prior is changing from one cycle to the next as it is recalculated from the latest ensemble.

\subsection{Variational data assimilation}
\label{sec:Var}
Variational DA (3D- or 4D-Var) computes optimal state estimates by minimizing a loss function, but without the use of an ensemble.  This point is worth emphasizing: classical 3D and 4D-var are not ensemble methods and hence do not randomly sample a posterior distribution.  The statistical interpretation normally associated with variational methods is that they return only the mode of the posterior distribution (see \cite{HBM2016}).  %On the other hand, diffusion models are random sampling methods.  Therefore, because variational DA is commonplace in NWP, we describe here 3D- and 4D-Var methods, but we emphasize that these methods do not directly relate to diffusion-based methods. 

The loss function of 3D-Var is 
\begin{equation}
	\mathcal{L}_\text{3D-Var}(\bx_k) =\frac{1}{2} \left\| \bR^{-1/2}\left(\by_k-\bH \bx_k\right)\right\|^2+
	\frac{1}{2} \left\| \bB^{-1/2}\left(\bx_k-\boldsymbol \mu_k\right)\right\|^2,
\end{equation}
where $\bR$ and $\bB$ are the observation error and background covariance matrices and where $\boldsymbol \mu_k$ is a background state (typically a forecast of some sort).

Similarly, 4D-Var is a smoother and as such uses the forward model $\mathcal{M}(\cdot)$ in~\eqref{eq:Model} to estimate the state $\bx_{k-1}$ \emph{before} the observation time $k$ by minimizing the loss function
\begin{equation}
	\mathcal{L}_\text{4D-Var}(\bx_{k-1}) =\frac{1}{2} \left\| \bR^{-1/2}\left(\by_k-\mathcal{H}(\mathcal{M}(\bx_{k-1})\right)\right\|^2+
	\frac{1}{2} \left\| \bB^{-1/2}\left(\bx_{k-1}-\boldsymbol \mu_{k-1}\right)\right\|^2.
\end{equation}

Classical 3D and 4D-Var use time-independent background matrices that represent the forecast error in
\emph{short-term} forecasts of the length of the cycling interval.
The celebrated NMC method
(i.e. \cite{parrishDerber1992}) was invented to make an estimate of precisely this quantity.
For example, in the linear, Gaussian system we use below to illustrate our main results we will see that the prior error covariance matrix converges to a steady-state by balancing the tendency of observations to reduce the error with the tendency of the forecast model to increase the error.  This steady-state prior covariance matrix is precisely the quantity that the NMC method is attempting to estimate using time-averaged statistics.
In nonlinear/non-Gaussian systems, the prior covariance matrix does not reach a steady-state, but in weakly nonlinear/non-Gaussian systems, for which synoptic-scale weather is likely applicable, the prior error covariance matrix wanders around an equilibrium that could be interpreted as this steady-state. In this case, the NMC method attempts to find this time averaged (steady-state) covariance matrix.

\section{Diffusion modeling and diffusion data assimilation}
\label{sec:Diffusion}
We briefly review the basics of diffusion modeling and how to use it for diffusion DA. We also briefly describe aspects of data-driven, ML\slash AI forecasting models and how these are related to our ideas.

\subsection{Diffusion modeling}
\label{sec:DiffusionModeling}
The goal in diffusion modeling is to construct a procedure to sample a random variable with probability density function (PDF), $p(\bx)$,
given a sufficiently large number of samples from that distribution in the form of a training dataset \citep{sd2015,ho2020}.  
The standard approach is to set up a forward process in the form of a simple stochastic differential equation (SDE) and to then reverse that process.
In the forward process, we start with a sample from $p(\bx)$ and sequentially add noise to the sample.
A neural network is trained on the samples (and the successively noisy versions of it) to enable a reverse process,
which takes in noise and then outputs  a sample from the PDF $p(\bx)$.

Following \cite{karras2022}, we use the stochastic differential equation (SDE)
\begin{equation}
	\label{forward}
	\text{d}\mathbf{u} = \sqrt{2t}\text{d}\beta_t,
\end{equation}         
where $\beta_t$ is a standard Brownian motion (Wiener process), as the forward process.
Its solution at time $t$ has the property that 
\begin{equation}
	\label{ut|u0}
	\mathbf{u}_t|\mathbf{u}_0 \sim N(\mathbf{u}_0, t^2\bI),
\end{equation}
which means the the forward process amounts to adding Gaussian noise to samples drawn from~$p(\bx)$. We emphasize that $t$ here is not a physical time, but a ``virtual'' time of the forward process. 

\cite{ANDERSON_SDE} constructs a reverse process that, when initialized with noise and integrated backwards in time from $t=T$ to $t=0$, generates samples from the PDF, $p(\bf{x})$, provided $T$ is large enough.
The reverse process is governed by the SDE
\begin{equation}
	\label{reverse1}
	\text{d}\mathbf{v} = -2t\nabla \log(p_t(\mathbf{v}))\text{d}t + \sqrt{2t}\text{d}\overline{\beta}_t
\end{equation}
where $p_t(\mathbf{v})$ is the time evolution of the marginal PDF of \eqref{forward} and $\overline{\beta}_t$ is a reverse-time Brownian motion.  
Thus, one can use the reverse process to generate samples from $p(\bx)$ by (i) drawing an initial condition from a Gaussian with mean zero and variance $T^2$ where $T$ is large; and (ii) simulating the reverse process \eqref{reverse1} from time $T$ to time $0$.

The term $\nabla \log p_t(\mathbf{v})$ is referred to as the ``score function'', which can be computed via Tweedie's formula (see Appendix~A and \cite{efron2011})
\begin{equation}
	\label{score}
	\nabla \log p_t(\mathbf{v}) = \frac{E_{\mathbf{x}\sim p(\mathbf{x}|\mathbf{v}_t)}[\bx]-\mathbf{v}_t}{t^2},
\end{equation}
where
\begin{equation}
	\label{eq:expectation}
	E_{\mathbf{x}\sim p(\mathbf{x}|\mathbf{v}_t)}[\mathbf{x}] = \int_{-\infty}^{\infty}\mathbf{x}p(\mathbf{x}|\mathbf{v}_t)\text{d}\mathbf{x}
\end{equation}
is a conditional expectation and, hence, the minimum variance estimator of $\bx$ given $\bv$. 
We can thus compute the conditional expectation as the minimizer of the loss function
\begin{equation}
	\ell = E_{\mathbf{x} \sim p(\mathbf{x})}E_{n\sim \mathcal{N}(0,t^2)}||\mathbf{x}-\mathbf{D}(\mathbf{v}=\mathbf{x}+n,t)||_2^2
\end{equation}
where $||\cdot||_2^2$ denotes the $\ell_2$ norm and where $D(\mathbf{v}_t, t)$ is referred to as a ``denoiser.''
In practice, the denoiser is a neural network that is trained to predict $\mathbf{x}$ from $\mathbf{v}_t$ as follows. 
For each sample $\mathbf{x}$, we have noisy versions at various times in the forward process,
which we can obtain by simply adding noise to the sample $\mathbf{x}$ according to the forward process.
These ``sample'' and ``noisy sample'' pairs are used to train the denoiser.  
Once trained, the denoiser is used for simulating the reverse process
\begin{equation}
	\label{reverseWithDenoiser}
	\text{d}\mathbf{v} = -2 \frac{D(\mathbf{v},t)-\mathbf{v}}{t}
	\text{d}t + \sqrt{2t}\text{d}\overline{\beta}_t,
\end{equation}
initialized with noise, backwards in time to sample the PDF $p(\bx)$.

\subsection{Diffusion data assimilation}
\label{sec:DiffusionDA}
A common extension of the diffusion modeling described just above is \emph{conditional} image generation in which one generates images of a requested scene given text prompts \citep[see, e.g.,][]{Zhan2024,ding2025}.
Conditional image generation via diffusion modeling is applicable to DA:
We train a diffusion model to take in noise and then generate (atmospheric) system states.
The only difference to the diffusion model outlined above is that we condition on the observations, $\by_k$, i.e., the denoiser is the minimizer of the loss function
\begin{equation}
	\label{eq:DenoiserConditioned}
	\ell = E_{\mathbf{x}_k,\by_k \sim p(\mathbf{x}_k,\by_k)}E_{n\sim \mathcal{N}(0,t^2)}||\mathbf{x}_k-\mathbf{D}(\mathbf{v}=\mathbf{x}_k+n,\by_k, t)||_2^2,
\end{equation}
where $p(\mathbf{x}_k,\by_k)$ is the \emph{joint} posterior. 
Conditioning is well-understood in the diffusion modeling literature \citep[see, e.g.,][]{batzolis2021, simpleObLike2024, Qu2024}.

Diffusion DA thus follows this recipe:
(i) we collect a large training dataset in the form of a time-series of system states $\bx_k$ and observations $\by_k$;
(ii) we set up a denoiser in the form of a neural network to minimize the loss function in \eqref{eq:DenoiserConditioned}.
This training is once and for all and (usually) never repeated.
The result is a diffusion model that takes in the latest observations and then generates an atmospheric system state.  
Examples of the use of diffusion models that work in this way include
\cite{Qu2024, pathak2024, li2025}. 
Other variants of diffusion DA supplement the observations $\by_k$ with an additional predictor in the form of a forecast,
either produced by the diffusion DA or by other means \citep[see, e.g.,][]{diffDA2024}.

One can also use \emph{approximations} to the denoiser that do not need to be explicitly trained on a particular observation network
\citep{simpleObLike2024, rozet2023, manshausen2024}.
The idea here is to think of the posterior PDF to be sampled from as
\begin{equation}
    p_t(\mathbf{v}|\mathbf{y}) \propto p_t(\mathbf{y}|\mathbf{v})p_t(\mathbf{v})
\end{equation}
where we emphasize that $\bv$ is a function of the time, $t$, of the diffusion process and we incorporate a subscript $t$ on the PDF here to emphasize that the PDF changes with the time of the diffusion process. Thus, as time in the diffusion process approaches zero, i.e., $t \rightarrow 0$, then $p_t(\mathbf{v}) \rightarrow p_0(\bv)$, which is the PDF describing the training set.
When using a long time series of past data for training, $p_0(\bv)$ is the ``climatological'' PDF.  
Perhaps more importantly, note further that $p_t(\mathbf{y}|\mathbf{v})$ approaches the known Gaussian observation likelihood also as $t \rightarrow 0$.  We use this fact next.  

The desired score function is  
\begin{equation}
    \label{scoreTrick}
    \nabla \log(p_t(\bv|\mathbf{y})) = \nabla \log(p_t(\mathbf{y}|\bv)) + \nabla \log(p_t(\bv))
\end{equation}
where the gradient is over the state, $\bv$.  The second term on the right is the unconditional score found in (\ref{score}) and determined from the climatological prior.  The first term on the right is approximated as
\begin{equation}
\label{eq:ApproxTrick}
    \nabla \log(p_t(\mathbf{y}|\bv)) \approx \nabla \log(p_0(\mathbf{y}|D(\bv,t)))
\end{equation}
where $p_0(\mathbf{y}|D) = p_{t\rightarrow 0}(\mathbf{y}|D)$, which as stated above is equal to the known Gaussian observation likelihood but conditioned on the result from the the \emph{unconditional} denoiser, $D(\bv,\tau)$, determined from the climatological prior. 
Assuming the observation likelihood has a covariance matrix, $\bR = r\bI$, equation (\ref{scoreTrick}) can be written as
\begin{equation}
    \label{eq:approxScore}
    \nabla \log(p_t(\bv|\mathbf{y})) \approx \frac{D(\bv,t) - \bv}{t^2} - \frac{1}{2r}\nabla||\by - \bH D(\bv,t)||_2^2  
\end{equation}

This approximation offers enormous computational savings as a new observation type can be assimilated without any re-training.  On the other hand, the approximation in~\eqref{eq:ApproxTrick} causes errors which we will explore further below.  

\subsection{Other forms of ML assisted data assimilation}
\label{sec:other}
There has been a flurry of recent activity using, for example, the ERA5 reanalysis for training ML-based \emph{forecast} models  \citep[see, e.g.,][]{gencast,graphcast,neuralGCM,seeds,mardani2025}.
In terms of probability distributions, these ML methods sample conditional distributions of the type 
$p(\bx_{k+T} \vert \bx_{k},\bx_{k-1})$, where $T$ is the desired forecast lead time and where we (arbitrarily) stopped the conditioning two time steps backwards in time (as, e.g., GenCast does, \cite{gencast}).
ML-based forecast models can replace or augment the physics-based forecast model in current DA systems, with some results reported by \cite{ASAW25}.
Our focus, however, is find out if and how generative AI may replace an \emph{entire} DA system, which is quite different from integrating data-driven forecast models into existing DA systems.

\section{Analysis of diffusion DA in a linear Gaussian system}
\label{sec:LinearExample}
We show how different training datasets used in diffusion DA lead to the sampling of different Bayesian posterior distributions. 
Our analysis uses a linear, stochastic dynamical system, because it allows us to perform calculations without approximations. 
For example, we can directly compute the various posterior distributions.
Moreover, a very similar problem setup has been used extensively to study ensemble DA systems and in particular the collapse of particle filters \citep[see, e.g.,][]{SBBA08, BBA08,BBL08,SBM15,MH23,HM23}.

\subsection{Problem setup}
We consider the linear SDE 
\begin{equation}
	\label{gauss_model}
	\text{d}\mathbf{x}=-\frac{1}{2}\mathbf{x}\text{d}s + \text{d}\beta_s,
\end{equation}
where $\bx$ is the $n_x$-dimensional system state,
$\beta_{s}$ is a Wiener process, and $s$ is physical time.
Observations $\by_k$ are collected $\Delta s$ time units apart via a linear observation operator, $\mathbf{H}$, generating $n_y$ observations
\begin{equation}
	\label{obs}
	\mathbf{y}_k=\mathbf{H}\mathbf{x}_k+\beps_k,
\end{equation} 
where $\beps_k$ is a draw from $N(\mathbf{0}, r\mathbf{I})$.
To keep the analysis simple, $\bH$ is composed of rows of the identity matrix, i.e.,
we observe $n_y$ components of $\bx$ directly.

The climatological prior for this problem is obtained from a long simulation of the dynamics~\eqref{gauss_model}.  
For a diffusion DA system, this long model run is used as the training set.
The climatological prior, however, can also be calculated analytically by solving the steady-state Fokker-Planck equation
\begin{equation}
	\nabla \circ\left(\frac{1}{2}\mathbf{x}p\right) + \frac{1}{2}\nabla^2p=0,
\end{equation}
with the boundary condition that the function $p(\mathbf{x})$ vanishes at $|\mathbf{x}|\rightarrow\infty$.
The solution is the standard Gaussian, i.e.,
\begin{equation}
	\label{climoPDF}
	p(\mathbf{x}) = \frac{1}{(2\pi)^{\frac{n_x}{2}}}\exp\left(-\frac{1}{2}\mathbf{x}^T\mathbf{x}\right).
\end{equation}
Thus, the training dataset are snapshots of a long time series of the solution of~\eqref{gauss_model} or, equivalently, draws from the standard Gaussian, i.e. a Gaussian with mean zero and covariance as the identity.

\subsection{Diffusion DA and the climatological prior}
\label{sec:climoDA}
We consider the Bayesian posterior distribution with a climatological prior.
This distribution is Gaussian and, hence, determined by the mean and (co-)variance,
which can be computed analytically.
We show that a diffusion DA that is trained with a long simulation, generates samples from this Bayesian posterior distribution.

\subsubsection{Bayesian posterior}
\label{sec:ClimoOpt}
The Bayesian posterior in this case is 
\begin{equation}
    p(\mathbf{x}_k|\mathbf{y}_k) \propto p(\mathbf{y}_k|\mathbf{x}_k)p(\mathbf{x}_k)
\end{equation}
where the prior is the climatological one in equation (\ref{climoPDF}) and the observation likelihood is
\begin{equation}
    \label{obLike}
    p(\mathbf{y}_k|\mathbf{x}_k) \propto \exp\left(-\frac{1}{2}(\mathbf{y}_k-\bH\mathbf{x})^T\mathbf{R}^{-1}(\mathbf{y}_k-\bH\mathbf{x})\right)
\end{equation}
where $\bR = r\bI_{n_y}$.
Thus, the posterior mean and covariance for this posterior distribution can readily be computed by, for example, multiplying (\ref{climoPDF}) with (\ref{obLike}) and subsequently completing the square within the argument of the exponential, to obtain
\begin{align}
	\overline{\mathbf{x}}^{a}_k &=  \mathbf{H}^T[\mathbf{H}\mathbf{H}^T+r\mathbf{I}_{n_y}]^{-1}\mathbf{y}_k,\\
	\mathbf{P}^a &= \mathbf{I}_{n_x} - \mathbf{H}^T[\mathbf{H}\mathbf{H}^T+r\mathbf{I}_{n_y}]^{-1}\mathbf{H}.
\end{align}
Since we assume that $\bH$ only contains a subset of the rows of the identity matrix, all matrices that appear above are diagonal (no correlation between state variables).
Thus, we can examine the posterior mean element-wise and consider only the diagonal elements of the posterior covariance matrix.
For an observed grid-point, we have 
\begin{align}
	\label{postMean}
	[\overline{\mathbf{x}}^{a}_k]^j &= \frac{1}{1+r}y,\\
	\label{postVar}
	[\bP^a]^{jj} & = 1 - \frac{1}{1+r} = \frac{r}{1+r}.
\end{align}
where $y=[\by_k]^j$ is shorthand notation for the $j^{th}$ element of the observation vector, $\by_k$.
Note that the posterior covariance is independent of the observations.

\subsubsection{Diffusion DA (exact denoiser)}
\label{sec:DiffusionDA_Climo}
To setup a diffusion DA system for this problem, we use the forward process (\ref{forward}) and thus integrate
\begin{equation}
	\label{reverse}
	d\mathbf{v} = -2t\nabla \log(p_t(\mathbf{v}\vert \mathbf{y}_k))\text{d}t + \sqrt{2t}\text{d}\overline{\beta}_t
\end{equation}
backward in virtual time (from $t=T$ to $t=0$).
We use Tweedie's formula
\begin{equation}
	\label{score1}
	\nabla \log p_t(\mathbf{v}\vert \mathbf{y}_k) = \frac{E_{\mathbf{x}_k\sim p(\mathbf{x}_k\vert\mathbf{v},\by_k)}[\bx_k]-\mathbf{v}}{t^2},
\end{equation}
to compute the score, but we avoid neural networks, and compute the conditional expectation (or denoiser) \emph{analytically}, which is equivalent to assuming a well-trained neural network.
Note that 
\begin{equation}
\label{eq:DiffusionConditional}
	p(\mathbf{x}_k\vert \mathbf{v},\by_k)\propto p(\mathbf{v},\by_k\vert \mathbf{x}_k)p(\mathbf{x}_k)
	\propto p(\by_k\vert \mathbf{x}_k)p(\mathbf{v}\vert \mathbf{x}_k)p(\mathbf{x}_k),
\end{equation}
because $\bv$ and $\by_k$ are independent, $p(\by_k\vert \mathbf{x}_k)$ is the familiar likelihood in equation (\ref{obLike}), and $p(\mathbf{x}_k)$ is the climatological prior (the long training dataset).
Due to our simple definition of the forward process, the reverse process is such that
\begin{equation}
\label{eq:ReverseV}
 \bv = \bx_k + \mathbf{n}, \quad \mathbf{n}\sim\mathcal{N}(0,t^2\bI),
\end{equation}
so that 
\begin{equation}	
	p(\mathbf{v}\vert \mathbf{x}_k)\propto \exp\left(-
	\frac{1}{2t^2}(\bv-\bx_k)^T(\bv-\bx_k)
	\right).
\end{equation}
The likelihood is defined by \eqref{obLike} and, therefore, putting everything together gives
\begin{equation}
	p(\mathbf{x}_k\vert \mathbf{v},\by_k)\propto \exp\left(-
	\frac{1}{2}(\hat{\by}-\hat{\bH}\bx_k)^T\hat{\bR}^{-1}(\hat{\by}-\hat{\bH}\bx_k)
	\right)
	\exp\left(-
	\frac{1}{2}\bx_k^T\bx_k
	\right),
\end{equation}
where
\begin{equation}
	\label{Reff}
	\hat{\mathbf{y}} = \begin{bmatrix}
		\mathbf{v} \\ \mathbf{y}_k 
	\end{bmatrix},
	\quad
	\hat{\mathbf{H}}= \begin{bmatrix}
		\mathbf{I}_{n_x}  \\ \mathbf{H}
	\end{bmatrix},
	\quad
	\hat{\bR}=\begin{bmatrix}
		t^2\mathbf{I}_{n_x} & \mathbf{0} \\
		\mathbf{0} & r\mathbf{I}_{n_o}
	\end{bmatrix}.
\end{equation}
Again, completing the square within the exponential finds the formula for the conditional mean
\begin{equation}
	\label{climoD}
	E_{\mathbf{x}_k\sim p(\mathbf{x}_k|\bv,\by_k)}[\bx_k]=  \hat{\mathbf{H}}^T[\hat{\mathbf{H}}\hat{\mathbf{H}}^T+\hat{\bR}]^{-1}\hat{\by}.
\end{equation}
The reverse process of the diffusion DA thus becomes
\begin{equation}
	\label{reverse2}
	\text{d}\mathbf{v} = -\frac{2}{t}\left[\hat{\mathbf{H}}^T[\hat{\mathbf{H}}\hat{\mathbf{H}}^T+\hat{\mathbf{R}}]^{-1}\hat{\mathbf{y}}-\mathbf{v}\right]\text{d}t + \sqrt{2t}\text{d}\overline{\beta}_t.
\end{equation}
Note that the ``effective'' observation error covariance $\hat{\bR}$ implies that, early in the reverse process ($t^2 \gg r$), the state is drawn towards the observations, $\mathbf{y}_k$,
but when $t^2 \ll r$,  the state is drawn towards $\mathbf{v}$.  
Similarly, note that the noise term vanishes as $t \rightarrow 0$ 
but the drift term takes on a greater significance.
These two effects ensure that the ensemble obtained from the diffusion model has the correct mean and variance.  

As before, the simple observation operator $\bH$ and the simple relation between $\bx_k$ and $\bv$ in~\eqref{eq:ReverseV} imply that all matrices in the reverse process~\eqref{reverse2} are diagonal.
We thus can consider one element of the vector $\mathbf{v}$, which we call $v$ for simplicity. 
For an observed element of $\mathbf{v}$ we have
\begin{equation}
	\label{reverse3}
	\text{d}v = -\frac{2}{t}\Big[\frac{\frac{r}{1+r}}{\frac{r}{1+r}+t^2}v+\frac{\frac{t^2}{1+r}}{\frac{r}{1+r}+t^2}y-v\Big]\text{d}t + \sqrt{2t}\text{d}\overline{\beta}_t.
\end{equation}

This equation can be solved analytically (see Appendix B):
\begin{equation}
	\begin{split}
		v(0) = \frac{\frac{r}{1+r}}{\frac{r}{1+r}+T^2}v(T) + \frac{r}{(1+r)^2}\frac{T^2}{\frac{r}{1+r}(\frac{r}{1+r}+T^2)}y + \frac{r}{1+r}\int_{T}^{0}\frac{\sqrt{2t}}{\frac{r}{1+r}+t^2}\text{d}\overline{\beta}_t.
	\end{split}
\end{equation}
Note that $v(0)$ is the Gaussian random variable of interest and we can compute its mean as
\begin{equation}
	\langle v(0)\rangle = \frac{\frac{r}{1+r}}{\frac{r}{1+r}+T^2}\langle v(T)\rangle +\frac{r}{(1+r)^2}\frac{T^2}{\frac{r}{1+r}(\frac{r}{1+r}+T^2)}y.
\end{equation}
In the limit $T\rightarrow\infty$ we find
\begin{equation}
    \label{exactMean}
	\langle v(0)\rangle_{T\rightarrow\infty} = \frac{1}{1+r}y,
\end{equation}
which is the posterior mean we obtained via the Kalman filter in~\eqref{postMean}.

Similarly, we can compute the variance
\begin{equation}
	\begin{split}
		\langle (v(0) - \langle v(0)\rangle)^2\rangle = \frac{(\frac{r}{1+r})^2}{(\frac{r}{1+r}+T^2)^2}(v(T)-\langle v(T)\rangle)^2 
		+ \Bigg\langle \Bigg(\frac{r}{1+r}\int_{T}^{0}\frac{\sqrt{2t}}{\frac{r}{1+r}+t^2}\text{d}\overline{\beta}_t\Bigg)^2 \Bigg\rangle 
	\end{split}
\end{equation}
where we have used the fact that the cross-term vanishes.
The stochastic integral on the right-hand side requires the use of the It$\hat{\textrm{o}}$ isometry, i.e.,
\begin{equation}
	\begin{split}
		\Bigg\langle \Bigg(\int_{T}^{0}\frac{\sqrt{2t}}{\frac{r}{1+r}+t^2}\text{d}\overline{\beta}_t\Bigg)^2 \Bigg\rangle = \Bigg\langle \Bigg(-\int_{0}^{T}\frac{\sqrt{2t}}{\frac{r}{1+r}+t^2}\text{d}\overline{\beta}_t\Bigg)^2 \Bigg\rangle \\= 2\int_{0}^{T}\frac{t}{(\frac{r}{1+r}+t^2)^2}\text{d}t = \frac{T^2}{\frac{r}{1+r}(\frac{r}{1+r}+T^2)}
	\end{split}
\end{equation}
Finally, we find the variance to be
\begin{equation}
	\begin{split}
		\langle (v(0) - \langle v(0)\rangle)^2\rangle = \frac{\frac{r^2}{(1+r)^2}}{(\frac{r}{1+r}+T^2)^2}T^2 + \frac{r}{1+r}\frac{T^2}{\frac{r}{1+r}+T^2}
	\end{split}
\end{equation} 
which in the limit as $T\rightarrow\infty$ becomes
\begin{equation}
        \label{exactVar}
	\begin{split}
		\langle (v(0) - \langle v(0)\rangle)^2\rangle_{T\rightarrow\infty} = \frac{r}{1+r},
	\end{split}
\end{equation}
which is the same variance as the Bayesian posterior using a climatological prior in \eqref{postVar}.

Hence, we have now shown that a diffusion DA system that is trained with a long time-series of system states and observations, samples a Bayesian posterior distribution with a climatological prior.

\subsubsection{Diffusion DA (approximate denoiser)}
We use the approximate score in equation (\ref{eq:approxScore}) to develop a diffusion model and then examine the resulting distributions to see how close they are to the Bayesian posterior using a climatological prior. By making use of the same procedure that determined equation (\ref{climoD}) we find that the \emph{unconditional} climatological denoiser is
\begin{equation}
	D =  \frac{1}{1+t^2}\mathbf{\bv}.
\end{equation}
Additionally, we have that
\begin{equation}
    \nabla||\by - \bH D||_2^2 = -\frac{2}{1+t^2}\bH^T\Big[\by - \frac{1}{1+t^2}\bH \mathbf{\bv}\Big],
\end{equation}
and combining these two equations with (\ref{reverse}) leads to the diffusion model
\begin{equation}
    d\mathbf{v} = -2 \frac{t}{1+t^2}\bv
	\text{d}t - \frac{2}{r}\frac{t}{1+t^2}\bH^T\Big[\by - \frac{1}{1+t^2}\bH \mathbf{\bv}\Big]\text{d}t + \sqrt{2t}\text{d}\overline{\beta}_t.
\end{equation}
As in the previous section with the exact denoiser we now examine an observed grid point to obtain the following scalar equation
\begin{equation}
    \label{approxSDE}
    dv = 2t \frac{1+r(1+t^2)}{r(1+t^2)^2}v
	\text{d}t - \frac{2}{r}\frac{t}{1+t^2}y\text{d}t + \sqrt{2t}\text{d}\overline{\beta}_t.
\end{equation}
The solution to this equation is derived in Appendix C and is repeated here
\begin{equation}
    v(0) = \frac{e^{-\frac{1}{r}\frac{T^2}{1+T^2}}}{1+T^2}v(T) + \Big[1 - e^{-\frac{T^2}{r(1+T^2)}}\Big]y +e^{-\frac{1}{r}}\int_T^0\sqrt{2t}\frac{e^{\frac{1}{r}\frac{1}{1+t^2}}}{1+t^2}\text{d}\overline{\beta}_t.
\end{equation}
The mean of the ensemble produced by this diffusion model is
\begin{equation}
    \langle v(0) \rangle = \frac{e^{-\frac{1}{r}\frac{T^2}{1+T^2}}}{1+T^2}\langle v(T) \rangle + \Big[1 - e^{-\frac{T^2}{r(1+T^2)}}\Big]y
\end{equation}
which in the limit as $T\rightarrow\infty$ is
\begin{equation}
    \label{approxMean}
    \langle v(0) \rangle = \Big[1 - e^{-\frac{1}{r}}\Big]y.
\end{equation}
Upon comparing (\ref{approxMean}) to the true mean (\ref{exactMean}) one finds that (\ref{approxMean}) is biased high, which implies that it draws too closely to the observations.

Similarly, we can compute the variance of the ensemble from the diffusion model as
\begin{equation}
	\begin{split}
		\langle (v(0) - \langle v(0)\rangle)^2\rangle = \frac{e^{-\frac{2}{r}\frac{T^2}{1+T^2}}}{(1+T^2)^2}(v(T)-\langle v(T)\rangle)^2 
		+ \Bigg\langle \Bigg(e^{-\frac{1}{r}}\int_T^0\sqrt{2t}\frac{e^{\frac{1}{r}\frac{1}{1+t^2}}}{1+t^2}\text{d}\overline{\beta}_t\Bigg)^2 \Bigg\rangle. 
	\end{split}
\end{equation}
By making use of the It$\hat{\textrm{o}}$ isometry we obtain
\begin{equation}
	\begin{split}
		\langle (v(0) - \langle v(0)\rangle)^2\rangle = \frac{e^{-\frac{2}{r}\frac{T^2}{1+T^2}}}{(1+T^2)^2}T^2 
		+ \frac{r}{2}\Big[1 - e^{-\frac{2}{r}\frac{T^2}{1+T^2}}\Big].
	\end{split}
\end{equation}
Lastly, evaluating this formula in the limit as $T\rightarrow\infty$ gives
\begin{equation}
    \label{approxVar}
	\begin{split}
		\langle (v(0) - \langle v(0)\rangle)^2\rangle = \frac{r}{2}\Big[1 - e^{-\frac{2}{r}}\Big].
	\end{split}
\end{equation}
A careful comparison with (\ref{exactVar}) reveals that the approximate denoiser leads to a variance that is too small.  

Hence, we have now shown that a diffusion model trained with a long time series of past data and with an approximate denoiser approximates random samples from the posterior with a climatological prior.
Due to the approximation in the denoiser, the posterior mean is biased too high and the posterior variance is biased too low. 

\subsection{Diffusion DA and a cycling prior}
\label{sec:cyclingDA}
We now consider a cycling prior that propagates information from one cycle to the next. 
Ensemble DA targets the corresponding Bayesian posterior distribution using this cycling prior, i.e. equation (\ref{iterated}).
We will show that a diffusion DA system that samples a Bayesian posterior distribution with a cycling prior requires re-training at each cycle. 
The reason is that the prior is equivalent to the training dataset for the diffusion model and therefore the training dataset must be regenerated at each cycle consistent with the latest observations and the ``flow-of-the-day".

\subsubsection{Bayesian posterior}
\label{sec:IteratedBayesKalman}
The Bayesian posterior of interest here is (\ref{iterated}).  We compute the posterior means and variances but in the interest of clarity we focus on the steady state,
i.e., $\bP^f \to \bP^f_\infty$, $\bP_a \to \bP^a_\infty$, $\bK\to \bK_\infty$ for large $k$ and, therefore, we have that
\begin{align}
	\bP^a_\infty &= (\bI-\bK_\infty \bH)\bP^f_\infty,\\
	\label{eq:KalmanGainSteady}
	\bK_\infty & = \bP^f_\infty\bH^T \left(\bH \bP^f_\infty\bH^T+r\bI\right)^{-1},
\end{align}
Because both the cycling prior and the observation likelihood are Gaussian for this problem these equations can be derived, like in the previous section, by completing the square or, for those familiar with the Kalman formalism, by comparison to the Kalman filter.  The steady state mentioned above is reached in linear, Gaussian systems when the observation operator, $\bH$, and the observation error covariance, $\bR$, matrices are constants in time.

The posterior mean does not converge to a steady state, but at a large time $k$ can be written as
\begin{equation}
	\overline{\bx}_{ak} =  \overline{\bx}_{fk} + \bK_\infty\left(\by_k-\bH  \overline{\bx}_{fk}\right),
\end{equation}
where $ \overline{\bx}_{fk}$ is the forecast mean, i.e., the posterior mean of the previous cycle evolved forward by the model~\eqref{gauss_model}.
Because we observe $n_y$ elements of $\bx_k$ directly, all matrices above are diagonal and it is sufficient to focus on one observed variable of the state vector.
An element of the Kalman gain corresponding to an observed system state is
\begin{equation}
	[\bK_\infty]_j = \frac{\alpha^j}{\alpha^j+r},
\end{equation}
which can be obtained from~\eqref{eq:KalmanGainSteady} by setting $\bH\to 1$, $\bP_\infty^f\to \alpha^j$ and $\bI\to 1$.
Similarly, we obtain the posterior mean and variance of an observed state variable as
\begin{align}
	\label{iterPostMean}
	[\overline{\bx}^{a}_k]^j &= [\overline{\mathbf{x}}_{fk}]^j + \frac{\alpha^j}{\alpha^j+r}([y_k]^j-[\overline{\mathbf{x}}_{fk}]^j),\\
	\label{Piter}
	[\bP^a_\infty]^{jj} &= \alpha^j-\frac{(\alpha^j)^2}{\alpha^j+r} = \frac{\alpha^j r}{\alpha^j+r}.
\end{align}
Because the data carry no information about unobserved grid points in our simplified setup, unobserved state variables are simply random draws from the cycling prior.

\subsubsection{Diffusion DA}
\label{sec:DiffusionDAWithRetrainingLinear}
We consider a diffusion DA system that is retrained at every cycle using a time evolving training dataset. 
We will show that there is a way to construct such a diffusion DA system that samples this Bayesian posterior distribution with a cycling prior, i.e., the posterior distribution typically targeted by ensemble DA (see \cite{BAO2024} for another example of such a diffusion DA system).

The procedure is as follows.
\begin{enumerate}
    \item Initially, we build a diffusion model as in Section~\ref{sec:LinearExample}.\ref{sec:climoDA}.\ref{sec:DiffusionDA_Climo} for the PDF $p(\mathbf{x}_0|\mathbf{y}_0)$.
    \item We use this diffusion model to randomly sample $p(\mathbf{x}_0|\mathbf{y}_0)$ to generate a large training set.
    \item We use the forecast model, which can be either a numerical solution to \eqref{gauss_model} or ML/AI based, to push each member of this set forward to the time of the next set of observations, $\by_1$.
    \item We then use these forecasts as the training dataset representing the cycling prior $p(\bx_1\vert \by_0)$ to build a new diffusion model that produces samples from $p(\bx_1|\by_1, \by_{0})\propto p(\by_1\vert \bx_1) p(\bx_1\vert \by_0)$ using the same methodology as in Section~\ref{sec:LinearExample}.\ref{sec:climoDA}.\ref{sec:DiffusionDA_Climo}.
    \item We then repeat steps 3 and 4 for each new cycle.
\end{enumerate}
We imagine we have done this up to the $k^{th}$ cycle and that the covariances have converged to their steady-state values -- which they will, assuming the diffusion is trained with sufficiently many samples.  

The reverse process is 
\begin{equation}
	\label{reverseCycling}
	d\mathbf{v} = -2t\nabla \log(p_t(\mathbf{v}\vert \mathbf{y}_k,\dots,\by_1))\text{d}t + \sqrt{2t}\text{d}\overline{\beta}_t,
\end{equation}
and we use Tweedie's formula to compute the score
\begin{equation}
	\nabla \log p_t(\mathbf{v}\vert \mathbf{y}_k,\dots\by_1) = \frac{E_{\mathbf{x}_k\sim p(\mathbf{x}_k\vert\mathbf{v},\by_k,\dots,\by_1)}[\bx_k]-\mathbf{v}}{t^2}.
\end{equation}
As before, we avoid neural networks and compute the conditional mean analytically, repeating the steps of Section~\ref{sec:LinearExample}.\ref{sec:climoDA}.\ref{sec:DiffusionDA_Climo},
but replacing the climatological prior in~\eqref{eq:DiffusionConditional} with the cycling prior $p(\bx_k\vert \by_{k-1},\dots,\by_1)$, i.e., the time-evolving (regenerated at each cycle) training dataset.
The result is
\begin{equation}
	\label{iteratedD}
	E_{\mathbf{x}_k \sim p(\mathbf{x}_k|\by_k,\by_{k-1},...)}[\mathbf{x}_k]=  \overline{\mathbf{x}}_{fk} + \bP^f_\infty\hat{\mathbf{H}}^T[\hat{\mathbf{H}}\bP^f_\infty\hat{\mathbf{H}}^T+\hat{\bR}]^{-1}\hat{\mathbf{y}}
\end{equation}
with $\hat{\bH}$ and $\hat{\bR}$ as before, but where we now have
\begin{equation}
	\hat{\mathbf{y}} = \begin{bmatrix}
		\mathbf{v}-\overline{\mathbf{x}}_{fk} \\ \mathbf{y}-\mathbf{H}\overline{\mathbf{x}}_{fk}
	\end{bmatrix},
\end{equation} 
due to the cycling prior.
Plugging the result into the reverse process yields
\begin{equation}
	\label{diffyIter}
	\text{d}\mathbf{v} =  -\frac{2}{t}\left[\bP^f_\infty\hat{\mathbf{H}}^T[\hat{\mathbf{H}}\bP^f_\infty\hat{\mathbf{H}}^T+\hat{\mathbf{R}}]^{-1}\hat{\mathbf{y}}-(\mathbf{v}-\overline{\mathbf{x}}_{fk})\right]\text{d}t + \sqrt{2t}\text{d}\overline{\beta}_{t},
\end{equation}
Due to our simple observation operator, all matrices above are diagonal and we can compute the reverse process element-wise to be
\begin{equation}
	\label{rev3}
	\text{d}v = -\frac{2}{t}\left[\frac{\frac{\alpha^jr}{\alpha^j+r}}{\frac{\alpha^jr}{\alpha^j+r}+t^2}(v-\overline{x})+\frac{\frac{\alpha^jt^2}{\alpha^j+r}}{\frac{\alpha^jr}{\alpha^j+r}+t^2}(y-\overline{x})-(v-\overline{x})\right]\text{d}t + \sqrt{2t}\text{d}\overline{\beta}_t
\end{equation}
where $y=[\by_k]^j$ and $\overline{x}=[\overline{\bx}_{fk}]^j$.
We can solve this SDE analytically (see Appendix C):
\begin{align}
	v(0) &= \overline{x} + \frac{\frac{\alpha^jr}{\alpha^j+r}}{\frac{\alpha^jr}{\alpha^j+r}+T^2}(v(T)-\overline{x}) + \frac{(\alpha^j)^2r}{(\alpha^j+r)^2}\frac{T^2}{\frac{\alpha^jr}{\alpha^j+r}(\frac{\alpha^jr}{\alpha^j+r}+T^2)}([\by_k]^j-\overline{x}) \\
	& + \frac{\alpha^jr}{\alpha^j+r}\int_{T}^{0}\frac{\sqrt{2t}}{\frac{\alpha^jr}{\alpha^j+r}+t^2}\text{d}\overline{\beta}_t, \nonumber
\end{align} 
Note that $v(0)$ is the Gaussian random variable of interest and we compute its mean
\begin{equation}
	\langle v(0)\rangle =\overline{x} + \frac{\frac{\alpha^jr}{\alpha^j+r}}{\frac{\alpha^jr}{\alpha^j+r}+T^2}(\langle v(T)\rangle - \overline{x}) +\frac{(\alpha^j)^2r}{(\alpha^j+r)^2}\frac{T^2}{\frac{\alpha^jr}{\alpha^j+r}(\frac{\alpha^jr}{\alpha^j+r}+T^2)}(y-\overline{x}),
\end{equation}
which, in the limit as $T\rightarrow\infty$, simplifies to
\begin{equation}
	\langle v(0)\rangle_{T\rightarrow\infty} = \overline{x} +\frac{\alpha^j}{\alpha^j+r}(y-\overline{x}).
\end{equation}
The above result is equal to the cycling posterior mean derived from the Bayesian posterior in~\eqref{iterPostMean}.  
Next, we compute the variance of $v(0)$
\begin{equation}
	\begin{split}
		\langle (v(0) - \langle v(0)\rangle)^2\rangle = \frac{(\frac{\alpha^jr}{\alpha^j+r})^2}{(\frac{\alpha^jr}{\alpha^j+r}+T^2)^2}\langle(v(T)-\langle v(T)\rangle)^2\rangle
		+ \Bigg\langle \Bigg(\frac{\alpha^jr}{\alpha^j+r}\int_{T}^{0}\frac{\sqrt{2t}}{\frac{\alpha^jr}{\alpha^j+r}+t^2}\text{d}\overline{\beta}_t\Bigg)^2 \Bigg\rangle 
	\end{split}
\end{equation}
where we have used the fact that the cross-term vanishes.
Upon making use of the It$\hat{\textrm{o}}$ isometry we find the variance to be
\begin{equation}
	\begin{split}
		\langle (v(0) - \langle v(0)\rangle)^2\rangle = \frac{\frac{(\alpha^j)^2r^2}{(\alpha^j+r)^2}T^2}{(\frac{\alpha^jr}{\alpha^j+r}+T^2)^2} + \frac{\alpha^jr}{\alpha^j+r}\frac{T^2}{\frac{\alpha^jr}{\alpha^j+r}+T^2}
	\end{split},
\end{equation} 
and in the limit $T\rightarrow\infty$ we find
\begin{equation}
	\langle (v(0) - \langle v(0)\rangle)^2\rangle_{T\rightarrow\infty} = \frac{\alpha^jr}{\alpha^j+r},
\end{equation}
which is the posterior variance we obtained the Bayesian posterior in~\eqref{Piter}.  

We have thus shown that a cycling prior corresponds to a time-evolving (regenerated at each cycle) training dataset for diffusion DA.
Put differently, a diffusion DA system that targets the same Bayesian posterior distribution as an ensemble DA requires re-training the denoiser at each cycle.

\subsection{Diffusion DA with an extended likelihood}
\label{sec:quasi-iterated}
We now consider diffusion DA systems that ingest observations and a forecast, denoted by $\mathbf{f}_k$.
The forecast propagates information from the past into the current DA cycle and we assume that it is derived from the posterior mean at the previous cycle $k-1$.
We subsequently treat the forecast as a ``pseudo'' observation,
which emulates one of many possible methods of supplementing a diffusion DA system with a forecast. 
We make our assumptions based on what we think is most reasonable at the time of writing. 

In terms of priors, likelihoods and posterior distributions, we now consider the state being conditioned on the observations and the forecast $\mathbf{f}_k$, i.e., the Bayesian posterior distribution of interest is
\begin{equation}
	\label{fixed3}
	p(\mathbf{x}_k|\mathbf{y}_k, \mathbf{f}_k) \propto 
	p(\mathbf{y}_k|\mathbf{x}_k)
	p(\mathbf{f}_k|\mathbf{x}_k)
	p(\mathbf{x}_k),
\end{equation}
where, $p(\mathbf{x}_k)$ is the climatological prior,
$p(\mathbf{y}_k|\mathbf{x}_k)$ is the usual Gaussian observation likelihood and
$p(\mathbf{f}_k|\mathbf{x}_k)$ is an ``extended'' likelihood to include the forecast $\mathbf{f}_k$,
which we assume is independent of the observation at time $k$.
The extended likelihood requires that we create a model for how the forecast is related to the ``truth,''.  As we demonstrate in Appendix E, this relationship is linear for our linear\slash Gaussian test problem and we assume that 
\begin{equation}
	\label{forecast_like}
	\mathbf{f}_k = a \bx_k + \boldsymbol \varepsilon_f,\quad 
	\boldsymbol \varepsilon_f\sim\mathcal{N}(\mathbf{0},r_f \bI),
\end{equation}
where $0 \le a\le 1$ and $r_f$ are scalars, which we determine later on.  

\subsubsection{Bayesian posterior}
\label{sec:qickKalman}
As before, one can either complete the square within equation (\ref{fixed3}) or make use of the Kalman filter formalism to compute the posterior mean and covariance explicitly.
Since the forecast is treated as an additional observation, we find
\begin{equation}
	\overline{\mathbf{x}}_a =  \mathbf{H}_e^T[\mathbf{H}_e\mathbf{H}_e^T+\mathbf{R}_e]^{-1}\mathbf{y}_e
\end{equation}
where
\begin{equation}
	\mathbf{y}_e = \begin{bmatrix}
		\mathbf{y}_k \\ \mathbf{f}_k
	\end{bmatrix}, \quad
	\bH_e = \begin{pmatrix}
		\bH\\
		a\bI
	\end{pmatrix},\quad
	\bR_e = \begin{pmatrix}
		r\bI & \bO \\
		\bO & r_f\bI
	\end{pmatrix}.
\end{equation}
Repeating the same calculation as before gives the elements of the posterior mean and posterior variance:
\begin{align}
	\label{postMeanQuick}
	[\bx^a_k]^j &= \frac{1}{1+r+a^2 \frac{r}{r_f}} \left( y + a\frac{r}{r_f} f\right),\\
	\label{postVarQuick}
	[\bP^a]^{jj} & =  \frac{r}{1+r+a^2\frac{r}{r_f}}.
\end{align}
where $[\mathbf{f}_k]^j=f$.
We note that as $r_f\to\infty$, that this posterior disregards the forecast and we recover the results from Section~\ref{sec:LinearExample}.\ref{sec:climoDA}.\ref{sec:ClimoOpt} as expected.

\subsubsection{Diffusion DA}
\label{sec:extendedDiffy}
Given observations and a forecast, the reverse process of a diffusion model becomes 
\begin{equation}
	\label{reverseExt}
	d\mathbf{v} = -2t\nabla \log(p_t(\mathbf{v}\vert \mathbf{y}_k,\mathbf{f}_k))\text{d}t + \sqrt{2t}\text{d}\overline{\beta}_t,
\end{equation}
and we use Tweedie's formula to compute the score
\begin{equation}
	\nabla \log p_t(\mathbf{v}\vert \mathbf{y}_k,\mathbf{f}_k) = \frac{E_{\mathbf{x}_k\sim p(\mathbf{x}_k\vert\mathbf{v},\by_k,\mathbf{f}_k)}[\bx_k]-\mathbf{v}}{t^2}.
\end{equation}
We again avoid neural networks and compute the conditional mean analytically, repeating the steps of Section~\ref{sec:LinearExample}.\ref{sec:climoDA}.\ref{sec:DiffusionDA_Climo}, but extending the likelihood by the forecast, so that
\begin{equation}
	\label{Reff2}
	\hat{\mathbf{y}} = \begin{bmatrix}
		\mathbf{v} \\ \mathbf{y}_k \\ \mathbf{f}_k
	\end{bmatrix},
	\quad
	\hat{\mathbf{H}}= \begin{bmatrix}
		\mathbf{I}_{n_x}  \\ \mathbf{H} \\ a \mathbf{I}_{n_x}
	\end{bmatrix},
	\quad
	\hat{\bR}=\begin{bmatrix}
		t^2\mathbf{I}_{n_x} & \mathbf{0}_{n_x\times n_o} & \mathbf{0}_{n_x}\\
		\mathbf{0}_{n_o\times n_x} & r\mathbf{I}_{n_o} & \mathbf{0}_{n_o\times n_x}\\
		\mathbf{0}_{n_x} & \mathbf{0}_{n_x\times n_o} & r_f \bI
	\end{bmatrix}.
\end{equation}
Repeating the same steps as in Section~\ref{sec:LinearExample}.\ref{sec:climoDA}.\ref{sec:DiffusionDA_Climo} but with $\hat{\bR}$, $\hat{\bH}$ and $\hat{\by}$ as above gives the the mean of the observed variable,
\begin{equation}
	\langle v(0)\rangle = \frac{\gamma}{\gamma+T^2}\langle v(T)\rangle + \frac{\gamma}{r}\frac{T^2}{\gamma+T^2}y +a\frac{\gamma}{r_f}\frac{T^2}{\gamma+T^2}f,
\end{equation}
where we introduced
\begin{equation}
	\gamma =  \frac{r}{1+r+a^2\frac{r}{r_f}},
\end{equation}
as a shorthand notation for the posterior variance.
For $T\to\infty$ we obtain
\begin{align}
	\lim_{T\to \infty}  \langle v(0)\rangle =& \frac{\gamma}{r} y +a \frac{\gamma}{r_f}f,
\end{align}
which upon rearrangement is the same as the Bayesian posterior result in~\eqref{postMeanQuick}.
Similarly, we find the variance of an observed quantity to be
\begin{equation}
	\langle (v(0) - \langle v(0)\rangle)^2\rangle = \frac{\gamma^2T^2}{(\gamma+T^2)^2} + \gamma\frac{T^2}{\gamma+T^2},
\end{equation} 
which, for $T\to\infty$ is equal to $\gamma$ which is the variance of the Bayesian posterior in~\eqref{postVarQuick}.

Hence, we have now shown that a diffusion model that is trained on a long time-series of system states of a dynamical system along with observations and forecasts of that dynamical system results in a diffusion DA system that samples a Bayesian posterior distribution constructed with an extended likelihood, and, possibly more importantly, all the while using the climatological prior.

\section{Numerical illustration}
\label{sec:Numerics}
We illustrate the variants of diffusion DA systems with a specific example, which also re-iterates the connections with variants of Bayesian posterior distributions.  
In the diffusion modeling literature, forward and reverse processes are derived in continuous time and subsequently discretized.
We follow the same approach and discretize forward and reverse processes using simple forward Euler methods.
We discretize the model~\eqref{gauss_model} with a similar forward Euler scheme, but other methods, or even analytical solution, will yield similar results.

\subsection{Discretization of the dynamical model}
We solve the SDE in (\ref{gauss_model}) numerically using the forward Euler method, i.e.,
\begin{equation}
	\label{FE}
	\mathbf{x}_{k+1} = D\mathbf{x}_k + \sqrt{\Delta}\mathbf{w}_k,
\end{equation}
where $\Delta$ is the time step, $\mathbf{w}_k$ is a random draw from $N(\mathbf{0}, \mathbf{I}_{n_x})$, the state vector, $\mathbf{x}_k$, is of length $n_x=100$, and 
\begin{equation}
	D = 1 - \frac{\Delta}{2}.
\end{equation}

Because we are working with a linear, Gaussian system and a linear observation operator, the optimal DA system is the Kalman filter.
That is, we can propagate the mean and covariance matrix forward in time without recourse to an ensemble, i.e.,
\begin{align}
	\label{FEMean}
	\overline{\mathbf{x}}_{k+1} &= D\overline{\mathbf{x}}_k,\\
	\label{FECov}
	\mathbf{P}_{k+1}^f &= D^2\mathbf{P}_k^f + \Delta\mathbf{I}_{n_x}.
\end{align}
Because our forward Euler (FE) discretization is accurate to first-order in $\Delta$, equation (\ref{FECov}) converges at large time to
\begin{equation}
	\label{covBias}
	\mathbf{P}_c^f = \frac{4}{4-\Delta}\mathbf{I}
\end{equation}
rather than $\mathbf{P}_c=\mathbf{I}$ as expected from our continuous time analysis (see Equation~\eqref{climoPDF}).  
We will use this discrete time climatological covariance to ensure consistency in the experiments described below.  

\subsection{Diffusion DA implementation details}
We implement all three diffusion DA systems and discretize their reverse processes using a forward Euler scheme to integrate from $T=100$ to $t=0$; the time step is set to $0.1$.
At each cycle, we generate an ensemble of $10^4$ samples from which we compute the mean and variances.
Note that the diffusion DA system with a cycling prior re-uses the ensemble from the previous cycle for re-training.
The diffusion DA with an extended likelihood uses the posterior ensemble mean from the previous cycle to generate a single forecast, i.e. it does not propagate an ensemble forward in time.

The diffusion DA with an extended likelihood further requires that we specify the extended likelihood parameters $a$ and $r_f$.  
To find these, we use Kalman filtering as follows.
We initialize the process with $a=1$ and $r_f=1$ and cycle the Kalman filter with extended likelihood (Section~\ref{sec:LinearExample}.\ref{sec:quasi-iterated}.\ref{sec:qickKalman}) for $10^5$ cycles to collect truth and forecast pairs $([\bx_k]^j,[\mathbf{f}_k]^j)$ at each grid point $j$.
We then fit two functions to this large dataset of forecast-truth pairs.
First, we can fit a line through the data because the mean of $p(\mathbf{f}_k|\bx_k)$ is $a\bx_k$,
so that the slope of the line can be used as a candidate value for $a$.
Second, we can compute the MSE associated with the \emph{forecast}, i.e.,
\begin{equation}
	\label{forecastMSE}
	\text{MSE}_f = E\left[ ([\mathbf{f}_k]_j - [\bx_k]_j)^2\right] = (1-a)^2 [\bx_k]_j^2 +r_f,
\end{equation}
and subsequently fit a quadratic to obtain another estimate of the parameter $a$ and also an estimate of the parameter $r_f$.  We then repeat this process by re-running the cycling experiment with the new values of $a$ and $r_f$ until the two estimates of $a$ agree to two decimal places.  As an example, for the case of $\Delta = 0.1$ time units between observations, the system settled on $a=0.61$ and $r_f = 0.34$
and the time-averaged MSE of the Kalman filter was then very close to the posterior variance prediction.  Lastly, we emphasize that this complicated iterative process for obtaining these two parameters is only necessary here because we avoided neural networks for the denoiser.  If we had used a neural network, rather than analytic formulas, a sufficiently expressive model would have learned these parameters from the data.       

\subsubsection{Results}
\begin{figure}[tb]
	\centering
	\includegraphics[width=0.6\textwidth]{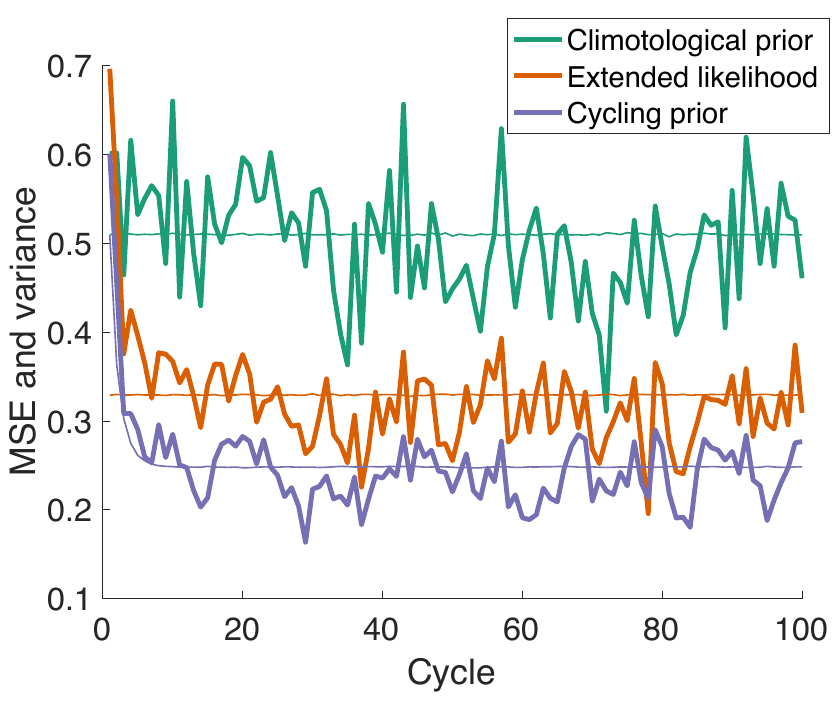}
	\caption{MSE (thick) and variance (thin) for diffusion DA systems with a climatological prior (green), an extended likelihood (orange) and a cycling prior (purple).
		MSE and variances from the Bayesian solutions nearly coincide with those of the diffusion DA systems, indicating that the two methods produce nearly identical results. }
	\label{fig:gaussMSE}
\end{figure}

We first run one experiment to confirm our theory.
Here, we observe every state element ($\mathbf{H}=\mathbf{I}$) every $\Delta=0.1$ time units and choose the observation error variance $r=1$. 
The posterior MSE and posterior variances are summarized in Figure~\ref{fig:gaussMSE},
for the three diffusion DA systems (climatological prior, extended likelihood and cycling prior).
As expected from our theory, we observe that (i) a diffusion DA system with a cycling prior generates the smallest MSE and posterior variance, while DA systems with a climatological prior or an extended likelihood lead to larger errors; (ii) bringing in a forecast via an extended likelihood improves the state estimates when compared to a DA system with only a climatological prior, but the errors and variances are still larger than what a fully cycled DA system can achieve; (iii) posterior variances match time-averaged MSE for all three diffusion DA systems, which is an important indicator that our theory is working correctly.
Thus, our example provides a numerical confirmation of our theory that the diffusion models with three different training datasets indeed realize three different Bayesian posterior distributions.

\begin{figure}[tb]
	\centering
	\includegraphics[width=1\textwidth]{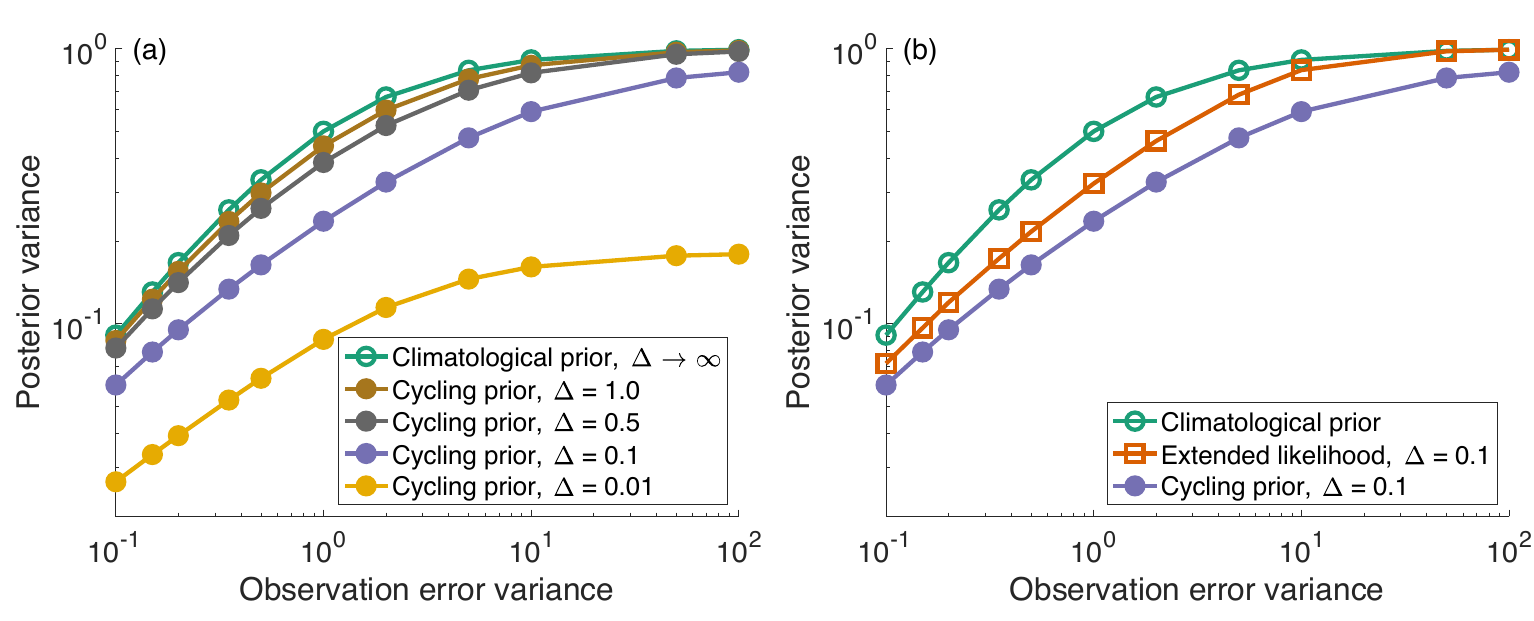}
	\caption{
		(a) Steady state posterior variance of a DA system with a cycling prior as a function of observation error and time interval between observations. 
		Also shown is the steady state posterior variance of a DA system with a climatological prior, which corresponds to a very large time interval between observations.
		(b) Steady state posterior variance of a DA system with an extended likelihood as a function of observation error (time interval between observations is $\Delta = 0.1$).
		Also shown are the posterior variances of a DA system with a cycling prior and with a climatological prior.}
	\label{fig:postVar}
\end{figure}

We perform additional numerical experiments in which we vary the time between observations $\Delta$ or the observation error $r$.
Intuitively, as the time interval between observations, $\Delta$, becomes larger, past observations carry less information and thus one would expect that the cycling prior converges to the climatological prior as $\Delta$ becomes large.  
We test this idea for various choices of $r$, with results summarized in Figure~\ref{fig:postVar}(a).
As the time between observations increases from $\Delta = 0.01$ to $\Delta = 2$, the posterior variance corresponding to a cycling prior converges to the posterior variance corresponding to a climatological prior (independently of $r$).  
Hence, the time between observations strongly controls the difference between those two forms of Bayesian posterior distributions and the resulting DA systems. 
More specifically, there is a delicate balance between the timescale of the error growth induced by the stochastic term in (\ref{gauss_model}) and the variance reducing property of the assimilation of observations.
If the error growth dominates (large time interval between observations), DA systems with a cycling prior are nearly identical to DA systems with a climatological prior.
If the assimilation of observations is frequent (short time interval between observations), then the cycling prior propagates information from past DA cycles to the current ones and, therefore, reduces state estimation errors and posterior variances.

Next, we perform a set of experiments in which we vary the observation error variance for a DA system with an extended likelihood. 
Results for $\Delta = 0.1$ are summarized in Figure~\ref{fig:postVar}(b), where we show the posterior variance of an extended likelihood DA system as a function of the observation error variance.
For comparison, we also plot the posterior variance of a DA system with a climatological prior and with a cycling prior (already shown in Figure~\ref{fig:postVar}(a)).
The posterior variance of a DA system with an extended likelihood is in between that of a system with a cycling or climatological prior, unless $r$ is very large (see below). 
Thus, for moderate $r$, the extended likelihood indeed propagates information from previous assimilation steps via a single forecast.  
The posterior variance of the extended likelihood DA system, however, is larger than that of a DA system with a cycling prior, which indicates that more information is transmitted between cycles when the entire distribution is propagated via an ensemble (i.e. a time-evolving training set), rather than a single forecast.  
Moreover, the posterior variance of a DA system with an extended likelihood converges to the posterior variance of a DA system with a climatological prior as the observation error variance $r$ becomes large.  
This is due to the fact that the information from the forecast is only as good as the information in the posterior mean it is integrated from.  
As the observation error increases we find that $r_f$ also increases such that there is less information in the forecast and therefore the differences between the posterior using a climatological prior and the posterior using the extended likelihood become muted.  
Note, however, that there is still a large difference between the posterior variance using a cycling prior and the posterior variance using the climatological prior even for very large observation error variances.  
Hence, we again see that propagating the entire distribution leads to more accurate state estimates than propagating a single forecast.  

\begin{figure}[tb]
	\centering
	\includegraphics[width=1\textwidth]{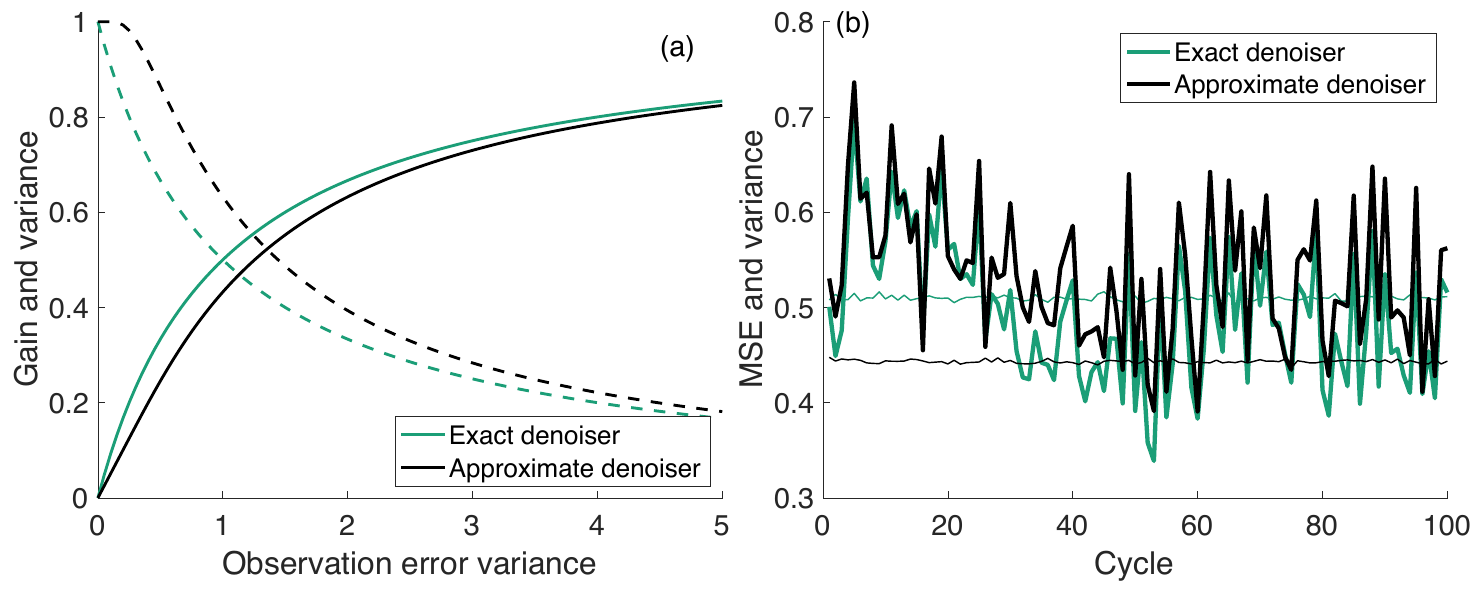}
	\caption{
		(a) Gain (dashed) and posterior variance (solid) for a diffusion DA system using a climatological prior as a function of observation error variance.
        Shown are the gain\slash variance of diffusion DA systems with climatological prior and an exact denoiser (green) or  approximate denoiser (black, see Section 4.b.3).  
		(b) MSE (thick) and variance (thin) for diffusion DA systems with a climatological prior (green) in a cycling DA experiment similar to Figure 1.  Here we compare the climatological diffusion DA system using the exact denoiser (green) to the approximate denoiser (black).}
	\label{fig:approxDenoiser}
\end{figure}

Lastly, we compare diffusion DA systems with a climatological prior using the exact and approximate denoisers (Section 4.b.3).  
In Figure \ref{fig:approxDenoiser}(a) we show the gain and posterior variance as a function of the observation error variance ($r$).  
Here, the ``gain'' is the quantity multiplying the observation when we compute the mean (compare to~\eqref{postMean}),
i.e., for the system with the exact denoiser, the gain is given by~\eqref{exactMean} and for the approximate denoiser the gain is~\eqref{approxMean}.
The variance of the system with the exact denoiser is given by~\eqref{exactVar} and the variance of the system with the approximate denoiser is~\eqref{approxVar}.
From this plot one can see that the gain of the system with the approximate denoiser is biased too large and therefore draws too closely to the observations.  This effect is strongest when the observation error is much smaller than the climatological error variance.  Similarly, the posterior variance produced by the approximate denoiser is too small.  These properties of the approximate denoiser transfer directly to explaining its behavior when used in a cycling DA system as shown in Figure \ref{fig:approxDenoiser}(b).  Here, we show MSE and variance as a function of DA cycle for diffusion DA systems with approximate and exact denoisers (both using climatological priors).
We see that the approximate denoiser has excess MSE as well as a posterior variance that is too small.  
Both of these properties are consistent with the biased gain and biased posterior variance. 

\section{Similarities and differences between diffusion DA, ensemble DA and variational DA}
Our analysis and numerical experiments reveal the relationship between the samples one receives from a particular diffusion DA system and the prior distribution used in training.  When one uses a climatological prior for training then that diffusion DA system is trained only once.  Additionally, we can use an approximate denoiser to avoid re-training when applying this DA system to new observational platforms, but the approximation in the denoiser results in unavoidable biases.
A diffusion DA system whose denoiser is retrained at every cycle samples a Bayesian posterior distribution with a cycling prior (the commonly used prior in ensemble DA). A diffusion DA system can be modified to ingest a forecast along with the observations. Such a system samples a Bayesian posterior distribution somewhere in between the one with a climatological and cycling prior.

Since diffusion DA and ensemble DA are sampling algorithms, connections between such systems are rather obvious. In particular, a diffusion DA system whose denoiser is re-trained at every cycle is equivalent to a traditional ensemble DA system, but it may be more robust to nonlinearity or non-Gaussianity. There is no traditional, equivalent ensemble DA system that replicates the diffusion DA system that samples a Bayesian posterior distribution with a climatological prior. The reason is likely that such a system is suboptimal, as our numerical examples show, because a cycling DA system achieves a smaller error and smaller variance due to the cycling prior. From an ensemble DA perspective, diffusion DA systems trained on a long dataset (climatological prior), are thus a rather unusual idea, but nevertheless these systems do seem to be a natural first choice for ML/AI modeling.  With that said, our analysis and numerical experiments suggest that such systems may not be able to achieve the accuracy of a DA system that uses a cycled prior.  We find the fact that diffusion DA systems with a climatological prior are theoretically sub-optimal, yet producing comparable results to cycled DA in practice, very interesting. 
One reason could be that the prior distributions found in NWP may have substantial non-Gaussianity, and diffusion DA methods may be able to account for non-Gaussianity more accurately than traditional ensemble DA.   

Diffusion models are sampling tools, while variational methods rely on optimization and only produce a single, optimal state estimate. Therefore, the connections between variational DA and diffusion modeling are somewhat murky, yet many papers on diffusion DA invoke similarities with 3D-Var or 4D-Var (e.g., \cite{manshausen2024}).
The focus of our paper is diffusion DA and the various Bayesian posterior distributions one can sample with these techniques and corresponding training datasets.  However, we can also comment on similarities and differences between diffusion DA and variational DA. One key difference here is how the background covariance matrix is constructed.
Traditional variational DA uses a constant-in-time background covariance. 
But this background covariance is \emph{not} the climatological covariance, but rather a covariance designed to represent short-term forecast errors (\cite{parrishDerber1992}). If one were to try to make a loose connection between variational and diffusion DA using a climatological prior, then one could argue that both methods use a static, time-independent prior or, equivalently, training dataset.
The training dataset used in diffusion DA (the climatological prior), however, does not correspond to the prior typically used in variational DA. Instead, the long training dataset used in this form of diffusion DA corresponds to an unusual and, as we have shown, suboptimal choice of the prior.

\section{Summary and conclusions}
\label{sec:Conclusions}
We have considered the question of how to use diffusion modeling to do data assimilation.
Specifically, we have worked out how different training datasets in diffusion modeling correspond to different prior assumptions and, hence, how these different diffusion DA systems sample from different Bayesian posterior distributions.
Our main results can be summarized as follows.
\begin{enumerate}
	\item 
	Diffusion DA is effective at sampling a Bayesian posterior distribution with a fixed, climatological prior, but the accuracy of such a system is inferior to an ensemble DA system that samples a Bayesian posterior distribution with a time-evolving, cycling prior that propagates information from one DA cycle to the next.
    Indeed, the training dataset in diffusion DA corresponds to a prior that is rather unusual in traditional DA.
	\item
	A diffusion DA system can sample the same posterior distribution as an ensemble DA system using a time-evolving cycling prior, but the denoiser in such a diffusion DA system will need to be re-trained at each DA cycle. 
	\item 
	A diffusion DA system with a fixed, climatological prior can be modified to ingest a forecast in addition to the observations.  
	This forecast adds some aspects of the time-evolving cycling prior back into the DA system, but is not entirely equivalent to a fully cycling ensemble DA. 
	If the forecast is generated by a separate DA system, the training cost is relatively low and the accuracy is in between the accuracy of a DA system with a climatological prior and a DA system with a cycling prior.
    \item 
    One can make use of an approximate denoiser when using the climatological prior.  While this results in a diffusion DA system that does not need re-training when making use of a new observation network, it does incur a penalty of a biased gain and a biased posterior variance.  Additionally, while it would be difficult to alleviate the bias in the gain, the posterior variance issue is likely to be easily fixed by the commonly available tunable parameters in the reverse diffusion process.  Lastly, because one already re-trains the diffusion model at every cycle when using the time-evolving, cycling prior the computational benefits of the approximate denoiser do not seem to apply.    
\end{enumerate}

Our results suggest that efficient implementation of diffusion DA systems with a cycling prior may yield better results than current diffusion DA systems. Possibilities for efficient training include fine-tuning methods \citep[see, e.g.,][]{parthasarathy2024}),  transfer learning \citep[see, e.g.,][]{zhuang2020}), or likelihood based approximations  \citep[see, e.g.,][]{simpleObLike2024}.  Work in these directions is already underway.  

\clearpage
%%%%%%%%%%%%%%%%%%%%%%%%%%%%%%%%%%%%%%%%%%%%%%%%%%%%%%%%%%%%%%%%%%%%%
% ACKNOWLEDGMENTS
%%%%%%%%%%%%%%%%%%%%%%%%%%%%%%%%%%%%%%%%%%%%%%%%%%%%%%%%%%%%%%%%%%%%%
\acknowledgments
DH is supported by the US Office of Naval Research (ONR) grant N0001422WX00451.
MM is supported by the US ONR grant N000142512298.

\vspace{2mm}\noindent
We thank Prof.~Ian Grooms, Prof.~Peter Jan van Leeuwen and an additional anomymous reviewer for their insightful comments that helped us improve the manuscript.

%%%%%%%%%%%%%%%%%%%%%%%%%%%%%%%%%%%%%%%%%%%%%%%%%%%%%%%%%%%%%%%%%%%%%
% DATA AVAILABILITY STATEMENT
%%%%%%%%%%%%%%%%%%%%%%%%%%%%%%%%%%%%%%%%%%%%%%%%%%%%%%%%%%%%%%%%%%%%%
% 
%
\datastatement
The code used for making the figures will be made available on github.

%%%%%%%%%%%%%%%%%%%%%%%%%%%%%%%%%%%%%%%%%%%%%%%%%%%%%%%%%%%%%%%%%%%%%
% APPENDIXES
%%%%%%%%%%%%%%%%%%%%%%%%%%%%%%%%%%%%%%%%%%%%%%%%%%%%%%%%%%%%%%%%%%%%%
%
\appendix[A]
\label{App:Tweedie}
	\appendixtitle{Tweedie's formula}
Here we derive Tweedie's formula specifically for the \emph{conditional} diffusion case where we have a conditional posterior distribution to sample from, but we emphasize that it is trivial to rework this derivation in the unconditional case by simply removing $y$ everywhere.  

Imagine we have a draw, $u$, from $p(u|y)$.  Additionally, imagine we generate a noisy version of this state, i.e.
\begin{equation}
	z = u|y + \epsilon
\end{equation}
where $\epsilon$ is a draw from $N(0, \sigma_o^2)$, which is assumed independent of $y$.  This implies that we can write
\begin{equation}
	p(z|y) = \int_{-\infty}^{\infty}p(z|u)p(u|y)du
\end{equation}
where
\begin{equation}
	p(z|u) = \frac{1}{\sqrt{2\pi \sigma_o^2}}exp\Big(-\frac{1}{2\sigma_o^2}(z-u)^2\Big)
\end{equation}
Similarly, we can find the derivative as
\begin{equation}
	\begin{split}
		\frac{d}{dz}(log(p(z|y))) = \frac{1}{p(z|y)}\frac{dp(z|y)}{dz} \\
		=-\int_{-\infty}^{\infty}\frac{(z-u)}{\sigma_o^2}\frac{p(z|u)p(u|y)}{p(z|y)}du \\
		= - \int_{-\infty}^{\infty}\frac{(z-u)}{\sigma_o^2}p(u|z,y)du \\
		= \frac{(-z+E[u|z,y])}{\sigma_o^2}
	\end{split}
\end{equation}
Therefore, we have that
\begin{equation}
	\label{tweedie}
	E[u|z,y] = z + \sigma_o^2\frac{d}{dz}(log(p(z|y))
\end{equation}
which is Tweedie's formula.
\appendix[B]
	\appendixtitle{Solution to the SDE in (\ref{reverse3})}
	We begin with (\ref{reverse3}) and perform a few simple manipulations:
	\begin{align}
		\label{rever3}
		\text{d}v &= -\frac{2}{t}\left[\frac{\frac{r}{1+r}}{\frac{r}{1+r}+t^2}v+\frac{\frac{t^2}{1+r}}{\frac{r}{1+r}+t^2}y-v\right]\text{d}t + \sqrt{2t}\text{d}\overline{\beta}_t,\\
		\label{rever4}
		\text{d}v &= -\frac{2}{t}\Big[-\frac{t^2}{\frac{r}{1+r}+t^2}v+\frac{\frac{t^2}{1+r}}{\frac{r}{1+r}+t^2}y\Big]\text{d}t + \sqrt{2t}\text{d}\overline{\beta}_t,\\
		\label{rever5}
		\text{d}v - 2t\frac{1}{\frac{r}{1+r}+t^2}v\text{d}t &= -2t\frac{\frac{1}{1+r}}{\frac{r}{1+r}+t^2}y\text{d}t + \sqrt{2t}\text{d}\overline{\beta}_t,\\
		\label{rever6}
		\frac{1}{\frac{r}{1+r}+t^2}\text{d}v - 2t\frac{1}{(\frac{r}{1+r}+t^2)^2}v\text{d}t &= -2t\frac{\frac{1}{1+r}}{(\frac{r}{1+r}+t^2)^2}y\text{d}t + \frac{\sqrt{2t}}{\frac{r}{1+r}+t^2}\text{d}\overline{\beta}_t.
	\end{align}
	\begin{equation}
		\label{rever7}
		d\Big(\frac{1}{\frac{r}{1+r}+t^2}v\Big) = -2t\frac{\frac{1}{1+r}}{(\frac{r}{1+r}+t^2)^2}y\text{d}t + \frac{\sqrt{2t}}{\frac{r}{1+r}+t^2}\text{d}\overline{\beta}_t.
	\end{equation} 
	Integrating backwards in time from $T$ to $0$ obtains
	\begin{equation}
		\begin{split}
			\int_{T}^{0}\text{d}\left(\frac{1}{\frac{r}{1+r}+t^2}v(t)\right)=-2y\int_{T}^{0}\frac{\frac{1}{1+r}t}{(\frac{r}{1+r}+t^2)^2}\text{d}t + \int_{T}^{0}\frac{\sqrt{2t}}{\frac{r}{1+r}+t^2}\text{d}\overline{\beta}_t,
		\end{split}
	\end{equation}
	leads to an expression for $v$ at time 0:
	\begin{equation}
		\label{eq:v0Early}
		\begin{split}
			v(0) = \frac{\frac{r}{1+r}}{\frac{r}{1+r}+T^2}v(T) -2y\frac{r}{(1+r)^2}\int_{T}^{0}\frac{t}{(\frac{r}{1+r}+t^2)^2}\text{d}t + \frac{r}{1+r}\int_{T}^{0}\frac{\sqrt{2t}}{\frac{r}{1+r}+t^2}\text{d}\overline{\beta}_t
		\end{split}.
	\end{equation}
	Using the change-of-variable, $s = \frac{r}{1+r}+t^2$, one can show that
	\begin{equation}
		\label{intRule}
		2\int_{T}^{0}\frac{t}{(\frac{r}{1+r}+t^2)^2}\text{d}t = -\frac{T^2}{\frac{r}{1+r}(\frac{r}{1+r}+T^2)}.
	\end{equation}
	Using this result in~\eqref{eq:v0Early} simplifies the expression to
	\begin{equation}
		\begin{split}
			v(0) = \frac{\frac{r}{1+r}}{\frac{r}{1+r}+T^2}v(T) + \frac{r}{(1+r)^2}\frac{T^2}{\frac{r}{1+r}(\frac{r}{1+r}+T^2)}y + \frac{r}{1+r}\int_{T}^{0}\frac{\sqrt{2t}}{\frac{r}{1+r}+t^2}\text{d}\overline{\beta}_t
		\end{split},
	\end{equation}
	which is the desired result.
\appendix[C]
        \appendixtitle{Solution to the SDE in (\ref{approxSDE})}
        We again begin by rewriting the SDE in (\ref{approxSDE}) as
        \begin{equation}
        dv - 2t \frac{1+r(1+t^2)}{r(1+t^2)^2}v
	   \text{d}t =- \frac{2}{r}\frac{t}{1+t^2}y\text{d}t +\sqrt{2t}\text{d}\overline{\beta}_t
        \end{equation}
        Next, by multiplying by an integrating factor we obtain
        \begin{equation}
        \frac{e^{\frac{1}{r}\frac{1}{1+t^2}}}{1+t^2}dv - 2t \frac{e^{\frac{1}{r}\frac{1}{1+t^2}}}{1+t^2}\frac{1+r(1+t^2)}{r(1+t^2)^2}v
	   \text{d}t =- \frac{e^{\frac{1}{r}\frac{1}{1+t^2}}}{1+t^2}\frac{2}{r}\frac{t}{1+t^2}y\text{d}t +\sqrt{2t}\frac{e^{\frac{1}{r}\frac{1}{1+t^2}}}{1+t^2}\text{d}\overline{\beta}_t
        \end{equation}
        \begin{equation}
        d\Bigg(\frac{e^{\frac{1}{r}\frac{1}{1+t^2}}}{1+t^2}v\Bigg) =- \frac{e^{\frac{1}{r}\frac{1}{1+t^2}}}{1+t^2}\frac{2}{r}\frac{t}{1+t^2}y\text{d}t +\sqrt{2t}\frac{e^{\frac{1}{r}\frac{1}{1+t^2}}}{1+t^2}\text{d}\overline{\beta}_t
        \end{equation}
        Integrating backwards in time from $T$ to $0$,
        \begin{equation}
        \int_T^0 d\Bigg(\frac{e^{\frac{1}{r}\frac{1}{1+t^2}}}{1+t^2}v\Bigg) =-\frac{2}{r} y\int_T^0 e^{\frac{1}{r}\frac{1}{1+t^2}}\frac{t}{(1+t^2)^2}\text{d}t +\int_T^0 \sqrt{2t}\frac{e^{\frac{1}{r}\frac{1}{1+t^2}}}{1+t^2}\text{d}\overline{\beta}_t
        \end{equation}
        obtains the following
        \begin{equation}
        e^{\frac{1}{r}}v(0) - \frac{e^{\frac{1}{r}\frac{1}{1+T^2}}}{1+T^2}v(T) =[e^\frac{1}{r} - e^\frac{1}{r(1+T^2)}]y +\int_T^0\sqrt{2t}\frac{e^{\frac{1}{r}\frac{1}{1+t^2}}}{1+t^2}\text{d}\overline{\beta}_t
        \end{equation}
        We now solve for $v(0)$ to obtain
        \begin{equation}
        v(0) = \frac{e^{-\frac{1}{r}\frac{T^2}{1+T^2}}}{1+T^2}v(T) + [1 - e^{-\frac{T^2}{r(1+T^2)}}]y +e^{-\frac{1}{r}}\int_T^0\sqrt{2t}\frac{e^{\frac{1}{r}\frac{1}{1+t^2}}}{1+t^2}\text{d}\overline{\beta}_t
        \end{equation}
        which is the desired result.
\appendix[D]
	\appendixtitle{Solution to the SDE in (\ref{rev3})}
	To solve the SDE
	\begin{equation}
		\label{reve3}
		\text{d}v = -\frac{2}{t}\Big[\frac{\frac{\alpha^jr}{\alpha^j+r}}{\frac{\alpha^jr}{\alpha^j+r}+t^2}(v-\overline{x})+\frac{\frac{\alpha^jt^2}{\alpha^j+r}}{\frac{\alpha^jr}{\alpha^j+r}+t^2}(y-\overline{x})-(v-\overline{x})\Big]\text{d}t + \sqrt{2t}\text{d}\overline{\beta}_t,
	\end{equation}
	we re-write it as
	\begin{equation}
		\label{rev4}
		\text{d}v = -\frac{2}{t}\Big[-\frac{t^2}{\frac{\alpha^jr}{\alpha^j+r}+t^2}(v-\overline{x})+\frac{\frac{\alpha^jt^2}{\alpha^j+r}}{\frac{\alpha^jr}{\alpha^j+r}+t^2}(y-\overline{x})\Big]\text{d}t + \sqrt{2t}\text{d}\overline{\beta}_t,
	\end{equation}
	and use the change of variables $v' = v - \overline{x}$ to obtain
	\begin{equation}
		\label{rev5}
		\text{d}v' - 2t\frac{1}{\frac{\alpha^jr}{\alpha^j+r}+t^2}v'\text{d}t = -2t\frac{\frac{\alpha^j}{\alpha^j+r}}{\frac{\alpha^jr}{\alpha^j+r}+t^2}(y-\overline{x})\text{d}t + \sqrt{2t}\text{d}\overline{\beta}_t.
	\end{equation}
	A few simple manipulations yield
	\begin{align}
		\label{rev6}
		\frac{1}{\frac{\alpha^jr}{\alpha^j+r}+t^2}\text{d}v' - 2t\frac{1}{(\frac{\alpha^jr}{\alpha^j+r}+t^2)^2}v'\text{d}t &= -2t\frac{\frac{\alpha^j}{\alpha^j+r}}{(\frac{\alpha^jr}{\alpha^j+r}+t^2)^2}(y-\overline{x})\text{d}t + \frac{\sqrt{2t}}{\frac{\alpha^jr}{\alpha^j+r}+t^2}\text{d}\overline{\beta}_t,\\
		\label{rev7}
		d\Big(\frac{1}{\frac{\alpha^jr}{\alpha^j+r}+t^2}v'\Big) &= -2t\frac{\frac{\alpha^j}{\alpha^j+r}}{(\frac{\alpha^jr}{\alpha^j+r}+t^2)^2}(y-\overline{x})\text{d}t + \frac{\sqrt{2t}}{\frac{\alpha^jr}{\alpha^j+r}+t^2}\text{d}\overline{\beta}_t.
	\end{align} 
	Integrating backwards in time, from $T$ to $0$,
	\begin{equation}
		\begin{split}
			\int_{T}^{0}d\Bigg(\frac{1}{\frac{\alpha^jr}{\alpha^j+r}+t^2}v'(t)\Bigg)=-2(y-\overline{x})\int_{T}^{0}\frac{\frac{\alpha^j}{\alpha^j+r}t}{(\frac{\alpha^jr}{\alpha^j+r}+t^2)^2}\text{d}t + \int_{T}^{0}\frac{\sqrt{2t}}{\frac{\alpha^jr}{\alpha^j+r}+t^2}\text{d}\overline{\beta}_t,
		\end{split}
	\end{equation}
	yields
	\begin{equation}
		\begin{split}
			v'(0) = \frac{\frac{\alpha^jr}{\alpha^j+r}}{\frac{\alpha^jr}{\alpha^j+r}+T^2}v'(T) -2(y-\overline{x})\frac{(\alpha^j)^2r}{(\alpha^j+r)^2}\int_{T}^{0}\frac{t}{(\frac{\alpha^jr}{\alpha^j+r}+t^2)^2}\text{d}t + \frac{\alpha^jr}{\alpha^j+r}\int_{T}^{0}\frac{\sqrt{2t}}{\frac{\alpha^jr}{\alpha^j+r}+t^2}\text{d}\overline{\beta}_t.
		\end{split}
	\end{equation}
	Making use of (\ref{intRule}) gives us
	\begin{equation}
		\begin{split}
			v'(0) = \frac{\frac{\alpha^jr}{\alpha^j+r}}{\frac{\alpha^jr}{\alpha^j+r}+T^2}v'(T) + \frac{(\alpha^j)^2r}{(\alpha^j+r)^2}\frac{T^2}{\frac{\alpha^jr}{\alpha^j+r}(\frac{\alpha^jr}{\alpha^j+r}+T^2)}(y-\overline{x}) + \frac{\alpha^jr}{\alpha^j+r}\int_{T}^{0}\frac{\sqrt{2t}}{\frac{\alpha^jr}{\alpha^j+r}+t^2}\text{d}\overline{\beta}_t.
		\end{split}
	\end{equation}       
	Undoing the change of variables finally yields
	\begin{equation}
		\begin{split}
			v(0) = \overline{x} + \frac{\frac{\alpha^jr}{\alpha^j+r}}{\frac{\alpha^jr}{\alpha^j+r}+T^2}(v(T)-\overline{x}) + \frac{(\alpha^j)^2r}{(\alpha^j+r)^2}\frac{T^2}{\frac{\alpha^jr}{\alpha^j+r}(\frac{\alpha^jr}{\alpha^j+r}+T^2)}(y-\overline{x}) + \frac{\alpha^jr}{\alpha^j+r}\int_{T}^{0}\frac{\sqrt{2t}}{\frac{\alpha^jr}{\alpha^j+r}+t^2}\text{d}\overline{\beta}_t,
		\end{split}
	\end{equation} 
	which is the desired result.
\appendix[E]
    \appendixtitle{Derivation of equation (\ref{forecast_like})}
    In the linear, Gaussian system of Section 4.a, the posterior mean using the climatological prior at time $k-1$ is a linear function of the observation, i.e.
\begin{equation}
    \label{3.1}
    \overline{\mathbf{x}}_a = \overline{\mathbf{x}}_c + \mathbf{K}_c[\mathbf{y}_{k-1}-\overline{\mathbf{x}}_c]
\end{equation}
Note further that the observation is a linear function of the truth of the form
\begin{equation}
    \label{3.2}
    \mathbf{y}_{k-1} = \mathbf{H}\mathbf{x}_{k-1} + \epsilon_{k-1}
\end{equation}
Lastly, note that the model to propagate the state from $k-1$ to $k$ would be
\begin{equation}
    \label{3.3}
    \mathbf{x}_k = D\mathbf{x}_{k-1} + \delta_{k-1}
\end{equation}
where we use the same numerical method as in Section 5 with
\begin{equation}
    \label{3.4}
    D = 1 - \frac{1}{2}\Delta
\end{equation}
and $\Delta$ is the time step of the model.
Setting the state at $k-1$ to the analysis in (\ref{3.1}) and then combining with (\ref{3.2}), (\ref{3.3}), and (\ref{3.4}) obtains
\begin{equation}
    \mathbf{f}_k = D\overline{\mathbf{x}}_c + D\mathbf{K}_c[\mathbf{H}\mathbf{x}_{k-1} + \epsilon_{k-1}-\overline{\mathbf{x}}_c] + \epsilon
\end{equation}
We now push the scalar factor $D$ through the bracket to obtain
\begin{equation}
    \mathbf{f}_k = D\overline{\mathbf{x}}_c + \mathbf{K}_c[\mathbf{H}D\mathbf{x}_{k-1} + D\epsilon_{k-1}-D\overline{\mathbf{x}}_c] + \epsilon
\end{equation}
Noting equation (\ref{3.3}) we can write this as
\begin{equation}
    \mathbf{f}_k = \mathbf{K}_c\mathbf{H}\mathbf{x}_k + D[\mathbf{I}-\mathbf{K}_c]\overline{\mathbf{x}}_c +\epsilon_{total}
\end{equation}
where $\epsilon_{total}$ is meant to represent the sum total of all the various noise terms.  We emphasize that this is clearly a linear function of the truth at time $k$.  Further, note that the Kalman gain in this setup has a diagonal that is less than one, which is why we pointed out that for this problem setup the parameter $a$ is bounded by $0 \le a \le 1$.  
    
%%%%%%%%%%%%%%%%%%%%%%%%%%%%%%%%%%%%%%%%%%%%%%%%%%%%%%%%%%%%%%%%%%%%%
% REFERENCES
%%%%%%%%%%%%%%%%%%%%%%%%%%%%%%%%%%%%%%%%%%%%%%%%%%%%%%%%%%%%%%%%%%%%%
% Make your BibTeX bibliography by using these commands:
\bibliographystyle{ametsocV6}
\bibliography{references}

\end{document}